\documentclass[aps,prb,twocolumn,amsmath,amssymb,superscriptaddress]{revtex4-2}
\usepackage{graphicx}
\usepackage{amssymb}
\usepackage{color}
\usepackage{amsmath}
\usepackage{float}
\usepackage{gensymb}
\usepackage{chemformula}
\usepackage{epstopdf}
\usepackage{hyperref}% add hypertext capabilities
\usepackage{makecell}
\usepackage{multirow}
\usepackage{afterpage}
\usepackage{tabularx}% Table width
\usepackage{amsmath}% Higher math
\usepackage{float}
\usepackage{booktabs}

\newcommand{\beq}{\begin{eqnarray}}
	\newcommand{\eeq}{\end{eqnarray}}

\hypersetup{hidelinks,
	colorlinks=true,
	allcolors=blue,
	pdfstartview=Fit,
	breaklinks=true}

\begin{document}
	\title{Anomalous Electrical Transport in the Kagome Magnet YbFe$_6$Ge$_6$
	}
	\author{Weiliang~Yao}
	\email{wy28@rice.edu}
	\affiliation{Department of Physics and Astronomy, Rice University, Houston, Texas 77005, USA}
	\affiliation{Department of Advanced Materials Science, The University of Tokyo, Kashiwa, Chiba 277-8561, Japan}
	\author{Supeng~Liu}
	\affiliation{Department of Advanced Materials Science, The University of Tokyo, Kashiwa, Chiba 277-8561, Japan}
	\author{Hodaka~Kikuchi}
	\affiliation{Institute for Solid State Physics, The University of Tokyo, Kashiwa, Chiba, 277-8581, Japan}
	\author{Hajime~Ishikawa}
	\affiliation{Institute for Solid State Physics, The University of Tokyo, Kashiwa, Chiba, 277-8581, Japan}
	\author{Øystein~S.~Fjellvåg}
	\affiliation{Laboratory for Neutron Scattering and Imaging, Paul Scherrer Institut, Villigen 5232, Switzerland}
	\affiliation{Department for Hydrogen Technology, Institute for Energy Technology, NO-2027 Kjeller, Norway}
	\author{David~W.~Tam}
	\affiliation{Laboratory for Neutron Scattering and Imaging, Paul Scherrer Institut, Villigen 5232, Switzerland}
	\author{Feng~Ye}
	\affiliation{Neutron Scattering Division, Oak Ridge National Laboratory, Oak Ridge, Tennessee 37831, USA}
	\author{Douglas~L.~Abernathy}
	\affiliation{Neutron Scattering Division, Oak Ridge National Laboratory, Oak Ridge, Tennessee 37831, USA}
	\author{George~D.~A.~Wood}
	\affiliation{ISIS Facility, STFC, Rutherford Appleton Laboratory, Chilton, Didcot, Oxfordshire OX11 0QX, United Kingdom}
	\author{Devashibhai~Adroja}
	\affiliation{ISIS Facility, STFC, Rutherford Appleton Laboratory, Chilton, Didcot, Oxfordshire OX11 0QX, United Kingdom}
	\affiliation{Highly Correlated Matter Research Group, Physics Department, University of Johannesburg, Auckland Park 2006, South Africa}
	\author{Chun-Ming~Wu}
	\affiliation{National Synchrotron Radiation Research Center, Hsinchu 30076, Taiwan}
	\author{Chien-Lung~Huang}
	\affiliation{Department of Physics and Center for Quantum Frontiers of Research \& Technology (QFort), National Cheng Kung University, Tainan 701, Taiwan}
	\author{Bin~Gao}
	\affiliation{Department of Physics and Astronomy, Rice University, Houston, Texas 77005, USA}
	\author{Yaofeng~Xie}
	\affiliation{Department of Physics and Astronomy, Rice University, Houston, Texas 77005, USA}
	\author{Yuxiang~Gao}
	\affiliation{Department of Physics and Astronomy, Rice University, Houston, Texas 77005, USA}
	\author{Karthik~Rao}
	\affiliation{Department of Physics and Astronomy, Rice University, Houston, Texas 77005, USA}
	\author{Emilia~Morosan}
	\affiliation{Department of Physics and Astronomy, Rice University, Houston, Texas 77005, USA}
	\affiliation{Smalley-Curl Institute, Rice University, Houston, Texas 77005, USA}
	\author{Koichi~Kindo}
	\affiliation{Institute for Solid State Physics, The University of Tokyo, Kashiwa, Chiba, 277-8581, Japan}
	\author{Takatsugu~Masuda}
	\affiliation{Institute for Solid State Physics, The University of Tokyo, Kashiwa, Chiba, 277-8581, Japan}
	\author{Kenichiro~Hashimoto}
	\affiliation{Department of Advanced Materials Science, The University of Tokyo, Kashiwa, Chiba 277-8561, Japan}
	\author{Takasada~Shibauchi}
	\email{shibauchi@k.u-tokyo.ac.jp}
	\affiliation{Department of Advanced Materials Science, The University of Tokyo, Kashiwa, Chiba 277-8561, Japan}
	\author{Pengcheng~Dai}
	\email{pdai@rice.edu}
	\affiliation{Department of Physics and Astronomy, Rice University, Houston, Texas 77005, USA}
	\affiliation{Smalley-Curl Institute, Rice University, Houston, Texas 77005, USA}
	
	\date{\today}
	
	\begin{abstract}
		Two-dimensional (2D) kagome metals offer a unique platform for exploring electron correlation phenomena derived from quantum many-body effects. Here, we report a combined study of electrical magnetotransport and neutron scattering on YbFe$_6$Ge$_6$, where the Fe moments in the 2D kagome layers exhibit an $A$-type collinear antiferromagnetic order below $T_{\rm{N}} \approx 500$ K. Interactions between the Fe ions in the layers and the localized Yb magnetic ions in between reorient the $c$-axis aligned Fe moments to the kagome plane below $T_{\rm{SR}} \approx 63$ K. Our magnetotransport measurements show an intriguing anomalous Hall effect (AHE) that emerges in the spin-reorientated collinear state, accompanied by the closing of the spin anisotropy gap as revealed from inelastic neutron scattering. The gapless spin excitations and the Yb-Fe interaction are able to support a dynamic scalar spin chirality, which explains the observed AHE. Therefore, our study demonstrates spin fluctuations may provide an additional scattering channel for the conduction electrons and give rise to AHE even in a collinear antiferromagnet.		
	\end{abstract}
	
	\maketitle
	
	The Hall effect, discovered by Edwin Hall, is the potential difference across an electric conductor transverse to the current and to an externally applied magnetic field perpendicular to the current \cite{Hall1879}. For conventional metals or semiconductors, the ordinary Hall effect (voltage) is linearly proportional to the applied magnetic field. The anomalous Hall effect (AHE), one of the most prominent phenomena of quantum transport in correlated electron materials, describes unconventional electron deflection other than the effect of external magnetic fields \cite{Hall1881,BergmannPhysicsToday1979,NagaosaRMP2010}. While the initial observation of AHE was made in ferromagnets (conventional AHE) \cite{Hall1881}, AHE can also occur in materials without net magnetization. The observation of AHE in antiferromagnets is of particular interest, where the interplay of itinerant electrons and spin texture (i.e., magnetic structure) gives rise to this unique transport property \cite{NakatsujiNature2015,ChenPRL2014,SmejkalNRM2022}. From the application perspective, antiferromagnetic (AFM) spintronic devices have advantages over ferromagnetic counterparts, such as the absence of stray fields and being robust to external magnetic field perturbation \cite{GomonayJLowTempPhys2014}.
	
	Therefore, a determination of the microscopic origin of the AHE in antiferromagnets has been attracting great interest. For example, it is well-known that a real-space Berry phase originating from a magnetic field-induced skyrmion lattice (non-coplanar spin texture) can produce an AHE termed ``topological Hall effect'' (THE) \cite{KanazawaPRL2011,KurumajiScience2019,HirschbergerNC2019,KimbellCommunMater2022,NagaosaNN2013,note}. The non-coplanar spin texture of a skyrmion has a non-zero scalar spin chirality (SSC) $\chi_{ijk} = \textbf{S}_i \cdot (\textbf{S}_j \times \textbf{S}_k)$, where $\textbf{S}_i$, $\textbf{S}_j$, and $\textbf{S}_k$ are three localized spins at sites $i$, $j$, and $k$, that acts as a fictitious magnetic field to induce AHE \cite{NagaosaRMP2010}. The chiral spin texture of a magnetic skyrmion is often stabilized by the competition of Heisenberg and Dzyaloshinskii-Moriya interactions \cite{KanazawaPRL2011,KurumajiScience2019,HirschbergerNC2019}.
	
	Going beyond the paradigm of the static spin effect, recent studies suggest that fluctuating spins can also contribute to the AHE by scattering the electrons, even in the spin-disordered state \cite{YangPRB2011,HouPRB2017,IshizukaSA2018,IshizukaPRB2021}. Arranging spins in a triangular motif is considered to be a favorable condition \cite{IshizukaPRB2021}, as spin fluctuations could be significantly promoted by the geometrical frustration. The kagome lattice, a two-dimensional (2D) network of corner-sharing triangles, is an ideal platform to realize this mechanism. Indeed, spin-fluctuation driven AHE/THE was recently observed in some kagome magnets, including $A$Mn$_6$Sn$_6$ ($A$ = Y, Sc, and Er) \cite{GhimireSA2020,WangPRB2021,ZhangAPL2022,FruhlingPRM2024}, $B_3$Ru$_4$Al$_{12}$ ($B$ = Nd and Gd) \cite{KolincioPNAS2021,KolincioPRL2023}, and HoAgGe \cite{RoychowdhuryPNAS2024}. In these materials, non-zero time-averaged SSC $\left\langle{\textbf{S}_i \cdot (\textbf{S}_j \times \textbf{S}_k)}\right\rangle$ is induced by spin fluctuations out of noncollinear magnetic structures or paramagnetic states. However, there is no direct evidence of such spin fluctuations, for example, by inelastic neutron scattering (INS) experiments, particularly in the temperature region where the AHE was detected. In addition, it is still unclear whether spin fluctuations from collinear AFM structures can have $\left\langle{\textbf{S}_i \cdot (\textbf{S}_j \times \textbf{S}_k)}\right\rangle \neq 0$ and therefore AHE/THE.
	
	\begin{figure}[t!]
		\hspace{-0.2cm}\includegraphics[width=8.0cm]{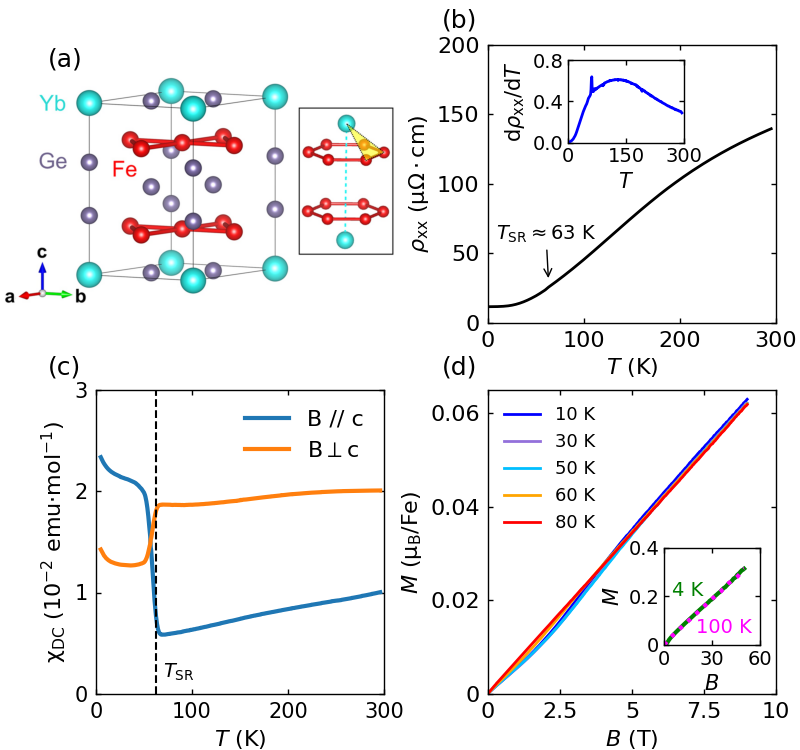}
		\caption{(a) Crystal structure of YbFe$_6$Ge$_6$. Right inset shows the local structure of Yb and Fe, with one spin triad of Yb-Fe highlighted in yellow. (b) Temperature dependence of zero-field resistivity in the kagome plane. Inset shows its derivative with respect to temperature. (c) Temperature dependence of DC magnetic susceptibility under a field of 1 T parallel and perpendicular to the $c$-axis. (d) Isothermal magnetizations along the $\textbf{a}^\ast$ direction up to 9 T at selected temperatures. Inset shows the magnetizations up to 50 T at 4 K and 100 K.}
		\label{fig1}
	\end{figure}
	
	Here we report the discovery of a spin-fluctuation driven AHE in the kagome lattice magnet YbFe$_6$Ge$_6$ with a collinear AFM structure \cite{VenturiniJAlloysCompd1992,MazetJPCM2000,AvilaJPCM2005,CadoganJPCM2009,RyanJPCS2010}. Through a comprehensive study of magnetization, magnetotransport, and neutron diffraction on single crystals, we find that the AHE emerges in its spin-reorientated state below $T_{\rm{SR}}\approx63$ K, where the collinear Fe spins rotate from the $c$-axis to the kagome plane. The combined space inversion and time-reversal ($\mathcal{IT}$) symmetry of the magnetic structure rules out the static non-zero SSC being the origin of the AHE \cite{NakatsujiNature2015,ChenPRL2014,SmejkalNRM2022}. Since our INS experiments reveal gapless low-energy spin excitations in the spin-reoriented state, the highly fluctuating Fe spins can contribute to a dynamic non-zero SSC by interacting with the Yb spins, which thereby induces the AHE \cite{WangNM2019}.
	
	The crystal structure of YbFe$_6$Ge$_6$ is shown in Fig. \ref{fig1}(a), where the Fe ions form a 2D kagome lattice. YbFe$_6$Ge$_6$ belongs to the family of hexagonal \textit{R}Fe$_6$Ge$_6$ compounds (with $R$ = Sc, Yb, Lu, Mg, Ti, Zr, Hf, and Nb) \cite{VenturiniJAlloysCompd1992,MazetSolidStateCommun2000,MazetJAlloysCompd2001,AvilaJPCM2005,MazetSolidStateCommun2013}. Their structure is derived by inserting the $R$ ions into the B35 structure FeGe \cite{VenturiniZeitschrift2006}. Despite the fact that they all have a N\'eel temperature well above room temperature, YbFe$_6$Ge$_6$ is the only member exhibiting a spontaneous spin reorientation (SR) transition \cite{VenturiniJAlloysCompd1992,MazetSolidStateCommun2000,MazetJAlloysCompd2001,AvilaJPCM2005,MazetSolidStateCommun2013}. Yb has been determined to be trivalent, yet its triangular lattice does not order at least above 0.4 K \cite{MazetJPCM2000,AvilaJPCM2005,SM}. Thus the magnetism of YbFe$_6$Ge$_6$ is mainly given by the Fe kagome lattice, although there must also be Yb-Fe interactions at low temperatures.
	
	The zero-field resistivity of YbFe$_6$Ge$_6$ along the kagome plane is presented in Fig. \ref{fig1}(b), which shows a typical metallic behavior with a residual-resistivity ratio of $\sim$12. The derivative of the resistivity shows a small yet sharp peak around $T_{\rm{SR}}$, indicating the change of magnetic structure. Fig. \ref{fig1}(c) shows the DC magnetic susceptibility under a magnetic field of 1 T along and perpendicular to the $c$-axis. The sudden susceptibility change around 63 K confirms the occurrence of the SR transition. These electrical and magnetic characteristics are consistent with an early report \cite{AvilaJPCM2005}.
	
\begin{figure}[t!]
		\hspace{-0.2cm}\includegraphics[width=8.5cm]{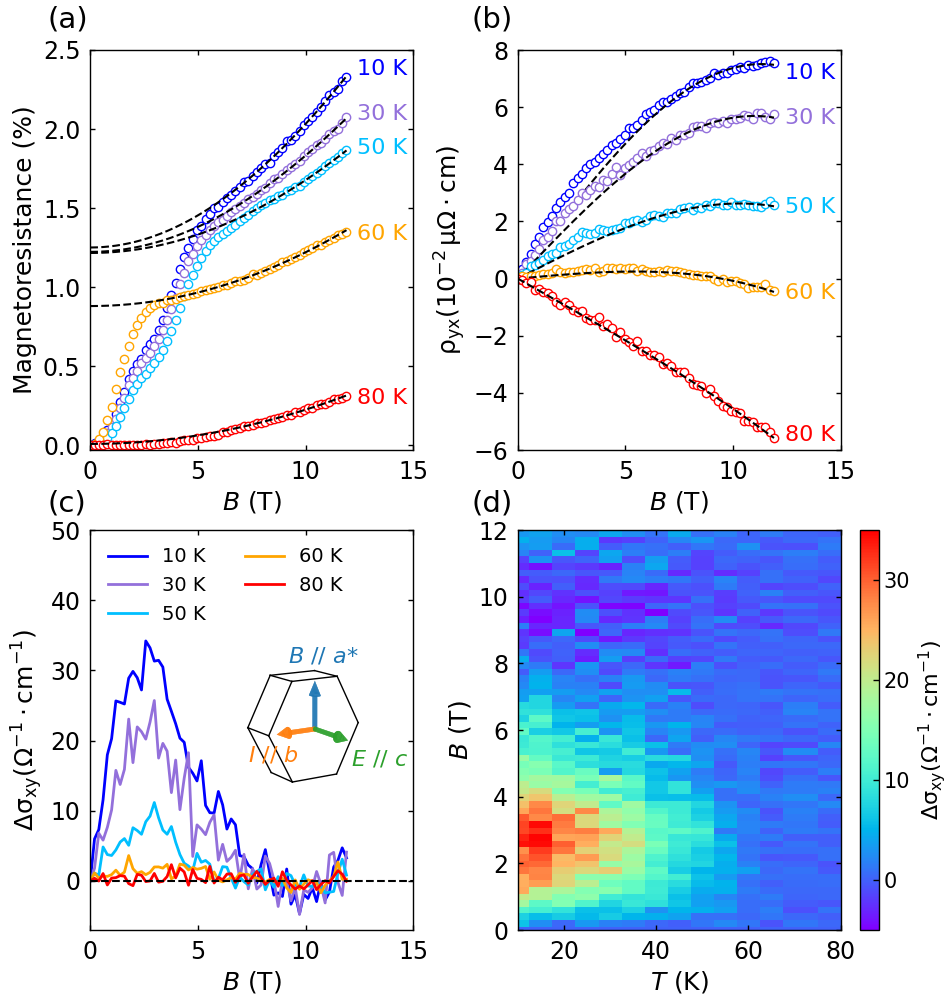}
		\caption{(a) Magnetoresistance at selected temperatures under fields along the $\textbf{a}^\ast$ direction. The electric current is along the $b$-axis. (b) Hall resistivity at selected temperatures under fields along the $\textbf{a}^\ast$ direction. Dashed curves in (a) and (b) are the fits with the two-band model \cite{SM}. (c) Remnant Hall conductivity ($\Delta\sigma_{xy}$) after subtracting the ordinary Hall contribution. Inset shows the Hall measurement configuration. (d) Contour plot of $\Delta\sigma_{xy}$ for temperatures from 10 K to 80 K and magnetic fields from 0 T to 12 T.}
		\label{fig2}
	\end{figure}

	Next, we explore how external fields affect the magnetic and transport properties of YbFe$_6$Ge$_6$. Fig. \ref{fig1}(d) presents the isothermal magnetization along the $\textbf{a}^\ast$ direction at selected temperatures from 10 K to 100 K. Although the magnetization shows an overall linear dependence on the field, it is slightly nonlinear at low fields for $T < {T_{\rm SR}}$. However, the magnitude of magnetization is quite small. Even at a magnetic field of 50 T, the moment size is only about 0.3 $\mu_{\rm{B}}$/Fe [inset of Fig.\ \ref{fig1}(d)], which is much smaller than the full moment size of $\sim$1.5 $\mu_{\rm{B}}$/Fe \cite{VenturiniJAlloysCompd1992,MazetJPCM2000,CadoganJPCM2009,RyanJPCS2010,SM}. The contrast between $T > {T_{\rm SR}}$ and $T < {T_{\rm SR}}$ can be more clearly discerned in the magnetoresistance (MR) [Fig. \ref{fig2}(a)], with MR = [$\rho_{{xx}}({B}) - \rho_{{xx}}(\rm{0\,T})$]/$\rho_{{xx}}(\rm{0\,T})$. Here the magnetic field is also applied along the $\textbf{a}^\ast$ direction. At 80 K, the MR has an approximately quadratic dependence on the field up to 12 T. When cooling below $T_{\rm{SR}}$, it first increases rapidly at low field and then shows the approximately quadratic behavior at high fields. Similar MR behavior was observed in Fe$_3$Sn$_2$, which is typical of metallic magnets with an SR transition \cite{KumarPRB2019,WangPRL2024}.
	
	Field dependence of the Hall resistivity is shown in Fig. \ref{fig2}(b), where we find a drastic change of $\rho_{{yx}}$ from 10 K to 80 K. In magnetic systems, the Hall resistivity can be decomposed into three parts: $\rho_{{yx}} = \rho^{\rm{O}}_{{yx}} + \rho^{\rm{A}}_{{yx}} + \Delta\rho_{{yx}}$ \cite{SuzukiNP2016,WangNM2019}, where $\rho^{\rm{O}}_{{yx}}$ is ordinary Hall resistivity, $\rho^{\rm{A}}_{{yx}}$ is conventional anomalous Hall resistivity due to net magnetization ($\rho^{\rm{A}}_{{yx}} \propto {M}$), and $\Delta\rho_{{yx}}$ is additional AHE contributions that cannot be included in the former two parts. For a magnetic system with localized moments, $\Delta\rho_{{yx}}$ usually comes from static and/or dynamic spin textures. In our case, $M$ is very small and shows barely temperature dependence [Fig. \ref{fig1}(d)], which is in contrast to the temperature evolution of $\rho_{{yx}}$ [Fig. \ref{fig2}(b)]. Therefore, $\rho^{\rm{A}}_{{yx}}$ is negligible, and we focus on the remaining two parts.
	
	The nonlinear field dependence of MR and $\rho_{{yx}}$ indicates a dominant multiband behavior. We use a two-band model to simultaneously fit the MR and $\rho_{{yx}}$ \cite{Ashcroft1976,WangPRL2024,ZhuJPCM2018}. Although the data at high fields can be well described by the two-band model, evident deviation from it at the low field region can be observed below $T_{\rm{SR}}$ [Fig. \ref{fig2}(a) and (b)], indicating a contribution from $\Delta\rho_{{yx}}$. Fig. \ref{fig2}(c) shows the corresponding Hall conductivity $\Delta\sigma_{{xy}}$, where it only appears below $T_{\rm{SR}}$. With more detailed measurements at intermediate temperatures \cite{SM}, we construct a contour plot of $\Delta\sigma_{{xy}}$, which shows a remarkable similarity to those metallic magnets with novel Hall effects \cite{SuzukiNP2016,KurumajiScience2019}. The maximum magnitude of the Hall conductivity at 10 K is about 30 $\Omega^{-1}\rm{cm}^{-1}$, comparable with those in previous ``anomalous Hall antiferromagnets" \cite{SmejkalNRM2022}. For magnetic fields applied perpendicular to the kagome plane, however, we do not observe such an AHE \cite{SM}, suggesting its intimate connection to the magnetic anisotropy as will be discussed below.
	
	\begin{figure}[t!]
		\hspace{-0.5cm}\includegraphics[width=7.8cm]{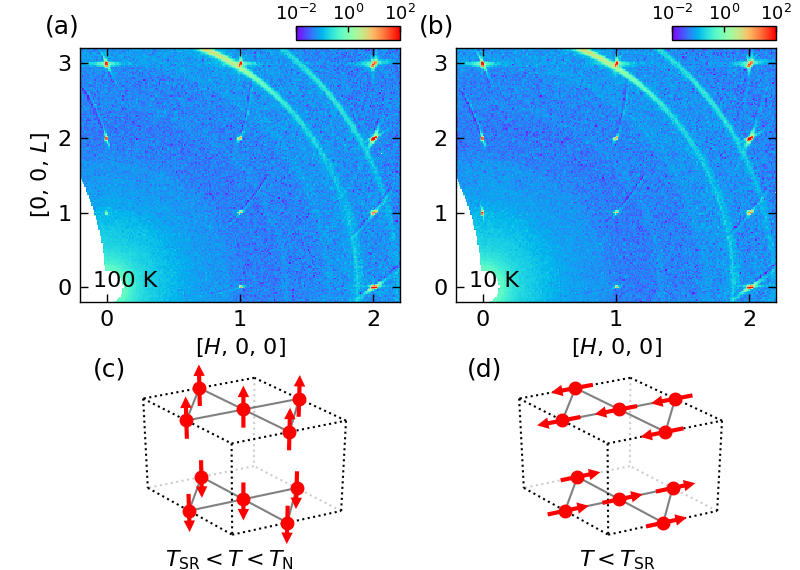}
		\caption{(a) and (b) Neutron diffraction patterns of the $(H, 0, L)$ plane at 100 K and 10 K, respectively. The intensity color bar is in log scale to highlight possible weak features. (c) and (d) Magnetic structures of the Fe kagome lattice above and below $T_{\rm{SR}}$, respectively.}
		\label{fig3}
	\end{figure}
	
	As the anomalous electrical transport is closely related to the SR, knowing the change of the magnetic structure across $T_{\rm{SR}}$ is crucial to uncovering its origin. We hence use single crystal neutron diffraction to study the magnetic structure of YbFe$_6$Ge$_6$. Fig. \ref{fig3}(a) shows the diffraction pattern of the $(H, 0, L)$ plane at 100 K ($> T_{\rm{SR}}$). There are no additional Bragg peaks except those at integer-index positions, consistent with the well-established $A$-type AFM structure with a propagation wave vector $\textbf{k}_{\rm{m}} = (0, 0, 0)$ [Fig. \ref{fig3}(c)] \cite{MazetJPCM2000,CadoganJPCM2009,RyanJPCS2010}. When cooling down to 10 K ($< T_{\rm{SR}}$), the diffraction pattern is strikingly similar [Fig. \ref{fig3}(b)] - no additional peaks appear, while the peak intensities have clear temperature-dependent behavior due to the SR \cite{SM}. This suggests the $\textbf{k}_{\rm{m}}$ of YbFe$_6$Ge$_6$ remains to be (0, 0, 0) below $T_{\rm{SR}}$. Representation analysis based on this propagation wave vector shows only a collinear structure with the spins along the $a$-axis [Fig. \ref{fig3}(d)] is compatible with the experiment. The SR transition in YbFe$_6$Ge$_6$ therefore equivalently flips the N\'eel vector by 90$^\circ$. More details of the magnetic structure analysis can be found in \cite{SM}. The magnetic structure below $T_{\rm{SR}}$ along with the lattice space group $P$6/$mmm$ ensures the system has the $\mathcal{IT}$ symmetry. With small-angle neutron scattering, we also confirmed no field-induced peaks appear under magnetic fields along the $\textbf{a}^\ast$ direction \cite{SM}. These observations rule out that the observed AHE is induced by static spin textures, such as magnetic skyrmions \cite{NagaosaNN2013}, which are restricted by Berry curvature that requires the absence of $\mathcal{IT}$ symmetry \cite{NagaosaRMP2010,ChenPRL2014,HouPRB2023}. Conversely, the spin-fluctuation driven AHE (or THE) is not necessarily dictated by this symmetry, as it originates from electron scattering by dynamic spins \cite{YangPRB2011,HouPRB2017,IshizukaSA2018,IshizukaPRB2021}.
	
	\begin{figure}[t!]
		\hspace{-0.3cm}\includegraphics[width=7.4cm]{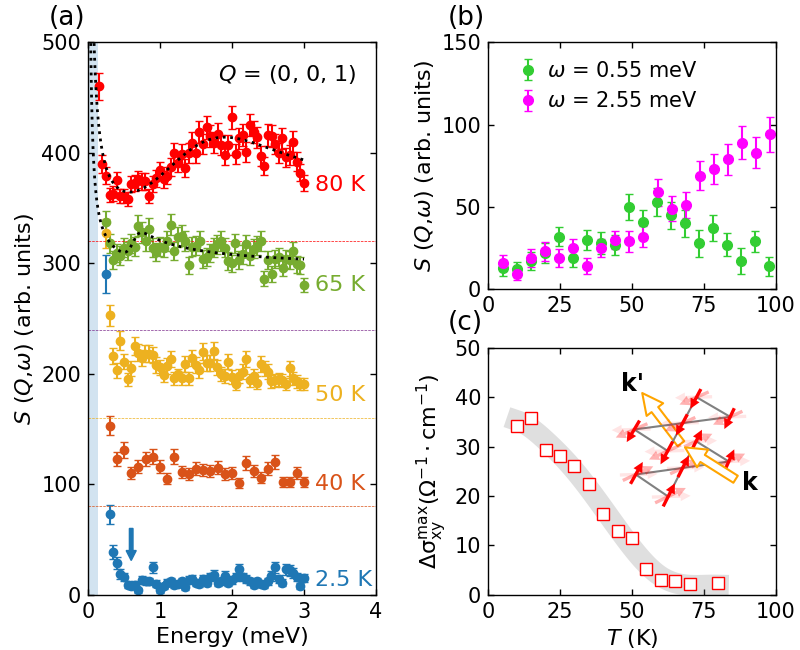}
		\caption{(a) Low-energy spin excitations at (0, 0, 1) at selected temperatures, offset for clarity. Horizontal dashed lines indicate the zero intensity for the  data above 2.5 K. Dashed curves are the fits to 65 K and 80 K data \cite{SM}. The light blue region shows energy resolution, and the arrow marks 0.6 meV, below which gapless excitations emerge. (b) Temperature dependence of the intensities at (0, 0, 1) at 0.55 meV and 2.55 meV. (c) Temperature dependence of the maximum magnitudes of $\Delta\sigma_{xy}$. The bold grey curve is a guide to the eyes. Inset illustrates electron scattering by the spin fluctuations.}
		\label{fig4}
	\end{figure}
	
	Since YbFe$_6$Ge$_6$ has localized magnetic moments, spin fluctuations below $T_{\rm{N}}$ are spin waves (magnons), which could impact the motion of conduction electrons \cite{YangPRB2011}. Hence, we use INS to directly probe the magnons and determine their temperature evolution. Fig. \ref{fig4}(a) shows the energy dependence of the INS intensities at (0, 0, 1). At 2.5 K, some INS intensities extend to $\sim$0.6 meV, which is much larger than our instrumental energy resolution ($\sim$0.14 meV). Therefore, the spin excitations are gapless in nature. As the temperature increases, their intensity significantly increases, but they remain gapless until 50 K. When the temperature goes beyond $T_{\rm{SR}}$, the intensity of the low-energy part starts to decrease. A broad peak centering around 2 meV shows up at 80 K, indicating the opening of a spin-wave gap. Such gap opening behavior suggests the magnetic anisotropy of YbFe$_6$Ge$_6$ transfers from an easy-plane type below $T_{\rm{SR}}$ to an easy-axis type above $T_{\rm{SR}}$, consistent with the magnetic structures presented in Fig. \ref{fig3}(c) and (d). By fitting the data with an error function multiplied by the Bose factor \cite{SM,ChenNC2024}, we can extract gap sizes of 0.64(4) meV and 1.34(5) meV at 65 K and 80 K, respectively.
	
	To determine the evolution of the spin-wave gap, we show the temperature dependence of the intensities at (0, 0, 1) at 0.55 meV and 2.55 meV [Fig. \ref{fig4}(b)], which are located below and above the spin-wave gap, respectively. Below $T_{\rm{SR}}$, the intensities at both energies increase as the temperature goes up. However, they start to bifurcate at $T_{\rm{SR}}$ - while the intensity above the spin-wave gap continues increasing, the intensity below it starts to decrease. The continuous increase of the intensity at 0.55 meV below $T_{\rm{SR}}$ further suggests it is an excitation signal rather than the tail of Bragg peak, whose intensity should otherwise decrease when the temperature approaches $T_{\rm{SR}}$. The onset of gapless spin excitations below $T_{\rm{SR}}$ aligns with the AHE [see Fig. \ref{fig4}(c)], suggesting their correlation.
	
	We point out that the gapless spin excitations below $T_{\rm{SR}}$ play a vital role in explaining the AHE through a dynamic SSC scattering mechanism, as first identified in the ferromagnetic heterostructure SrRuO$_3$/SrTiO$_3$ \cite{WangNM2019}. The gapless excitations at (0, 0, 1) within our energy resolution suggests that Fe spins can fluctuate at the cost of a very small energy, facilitating the formation of a dynamic SSC. From a semiclassical perspective, the propagation of spin waves causes adjacent Fe spins in the kagome plane to slightly deviate from the collinear ground state. On the other hand, Yb spins remain disordered above 0.4 K and thus, by themselves, have little impact on the AHE, which emerges at a much higher temperature. However, their interactions with Fe strengthen upon cooling, leading to the SR transition. Because the Yb positions and Fe kagome layers are well separated [Fig. \ref{fig1}(a)], the transient non-collinear Fe spin configurations and Yb spins can locally produce an SSC with $ \textbf{S}_i \cdot (\textbf{S}_j \times \textbf{S}_k) \neq 0$. Despite these local SSCs with opposite signs (+ or -) cancel out at zero magnetic field due to the global $\mathcal{IT}$ symmetry, a finite magnetic field partially aligns the Yb spins, resulting in a non-zero overall SSC \cite{SM}. This non-zero SSC can preferentially scatter electrons and give rise to the observed AHE. However, when the magnetic field becomes exceedingly large, low-energy spin fluctuations are suppressed due to the opening of a Zeeman gap, leading to a vanishing overall SSC and, consequently, a zero AHE. In contrast, the finite spin-wave gap above $T_{\rm{SR}}$ suggests relatively high spin stiffness. As a result, Fe spins favor maintaining the ground-state collinear configuration. Both the increased spin stiffness of the Fe kagome layer and the weaker Yb-Fe interaction at higher temperatures are unfavorable for the dynamic SSC, which accounts for the absence of the AHE above $T_{\rm{SR}}$.
	
	Therefore, the AHE in YbFe$_6$Ge$_6$ can be understood as arising from electron scattering by low-energy spin fluctuations. Given that every Fe ion has a static magnetic moment of approximately 1.5 $\mu_{\rm{B}}$ below $T_{\rm{SR}}$ \cite{SM}, we estimate a Zeeman gap of roughly 0.6 meV for a 7 T field \cite{TengNature2022,ChenNC2024}, beyond which the AHE vanishes at low temperatures [Fig. \ref{fig2}(c) and (d)]. It suggests the dynamic SSC contributing to the AHE is within 0.6 meV. In addition, this energy scale reasonably matches the gap size at 65 K [Fig. \ref{fig4}(a)], which is slightly above $T_{\rm{SR}}$ and the AHE just disappears [Fig. \ref{fig4}(c)]. For comparison, we note that the similar kagome antiferromagnet FeSn does not exhibit an AHE \cite{SalesPRM2019,KangNM2020}. Despite also having a collinear magnetic structure with spins lying in the kagome plane, the spin waves of FeSn display a finite energy gap of $\sim$2 meV \cite{XieCommunPhys2021,DoPRB2022}, which suppresses low-energy spin fluctuations.
	
	Recent experiments show that spin fluctuations can give rise to similar AHEs (or THEs) in a range of materials, including single crystals \cite{GhimireSA2020,WangPRB2021,KolincioPNAS2021,ZhangAPL2022,KolincioPRL2023,FruhlingPRM2024,RoychowdhuryPNAS2024,AbenpjQuantMats2024,JeonCommunPhys2024} and thin films \cite{WangNM2019,FujishiroNC2021}, supporting the idea that spin-fluctuation driven AHE is a general phenomenon. In previous studies, it depends on thermal fluctuations, with samples being heated above or close to the ordering temperature \cite{GhimireSA2020,WangPRB2021,KolincioPNAS2021,ZhangAPL2022,KolincioPRL2023,FruhlingPRM2024,RoychowdhuryPNAS2024,AbenpjQuantMats2024,JeonCommunPhys2024,WangNM2019,FujishiroNC2021}. Those AHEs are therefore essentially driven by paramagnetic spin fluctuations, which are usually dominant around magnetic Bragg peak positions near the transition temperature and are gapless as well \cite{Collins1967}. For YbFe$_6$Ge$_6$, the situation is different. Due to the SR transition induced by the Yb-Fe interaction, the $c$-axis aligned spins transfer to the easy-plane arrangement below $T_{\rm{SR}}$, which greatly promotes low-energy spin fluctuations even in the spin-ordered state. With the stronger Yb-Fe interaction at low temperatures, the AHE is enhanced accordingly. Moreover, YbFe$_6$Ge$_6$ adopts a collinear AFM structure that is much simpler than the magnetic structures of previous single crystals \cite{GhimireSA2020,WangPRB2021,KolincioPNAS2021,ZhangAPL2022,KolincioPRL2023,FruhlingPRM2024,RoychowdhuryPNAS2024,AbenpjQuantMats2024,JeonCommunPhys2024}. Our study therefore also suggests that a complex (e.g., chiral) static magnetic structure is not a prerequisite for observing spin-fluctuation driven AHE.
	
	In conclusion, with electrical transport, magnetization, and neutron scattering, we have systematically studied the Fe-based kagome magnet YbFe$_6$Ge$_6$. An AHE is observed in its AFM-ordered state when the spins are aligned parallel to the kagome plane due to the Yb-Fe interaction. The collinear magnetic structure with $\mathcal{IT}$ symmetry excludes a static SSC origin of the AHE. By directly measuring the low-energy spin excitations, we find that spin excitations at the Brillouin zone center become gapless in the spin-reorientated state. The simultaneous onset of these gapless excitations and AHE shows their close relationship and points to a dynamic SSC scattering mechanism of the latter. Our results not only reveal a spin-fluctuation driven AHE that is not solely dependent on thermal activation but also demonstrate dynamic spins, even in a collinear antiferromagnet, can significantly impact electrical transport.
	
	\begin{acknowledgments}
		The neutron scattering and single-crystal synthesis work at Rice was supported by US NSF-DMR-2401084 and by the Robert A. Welch Foundation under grant no. C-1839 (P.D.), respectively. The transport work at University of Tokyo was supported by Grant-in-Aid for Scientific Research (KAKENHI) (JP22H00105), and Grant-in-Aid for Scientific Research on Innovative Areas ``Quantum Liquid Crystals'' (JP19H05824) from Japan Society for the Promotion of Science. W.Y. acknowledges support from the US NSF-OISE-2201516 under the Accelnet program of Office of International Science and Engineering (OISE). P.D. also acknowledges travel support from the US NSF-OISE-2201516. A portion of this research used resources at the Spallation Neutron Source, a DOE Office of Science User Facility operated by the Oak Ridge National Laboratory. The experiment at the ISIS Neutron and Muon Source was supported by a beam time allocation RB2410134 from the Science and Technology Facilities Council. Ø.S.F. acknowledges funding from The Research Council of Norway through project 325345. This work is in part based on work performed at the DMC instrument at SINQ at the Paul Scherrer Institute. Small-angle neutron scattering was performed at QUOKKA, ANSTO under a proposal number P18792. CLH acknowledges the support by the National Science and Technology Council in Taiwan (NSTC 113-2112-M-006-027).
	\end{acknowledgments}	
	
	%%Bibliography
	\bibliographystyle{apsrev4-1}
	\bibliography{yfg_reference}
	
	%\centering{\textbf{End Matter}}
	\begin{center} \textbf{End Matter} \end{center}
	
\begin{figure}[h!]
		\hspace{-0.1cm}\includegraphics[width=8.8cm]{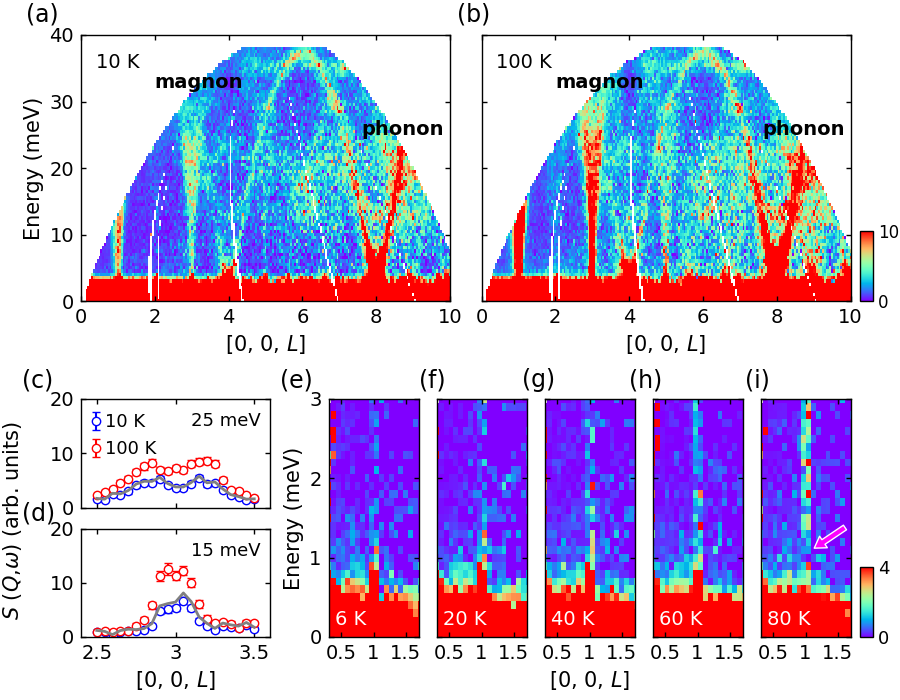}
		\caption{(a) and (b) Excitation spectra along [0, 0, $L$] direction at 10 K and 100 K, respectively. (c) and (d) Constant energy cuts around (0, 0, 3) at 25 meV and 15 meV, respectively. Grey curves represent the calculated intensities at 100 K by correcting the 10 K data with the Bose factor. (e)-(i) Low-energy spin excitations around (0, 0, 1) at selected temperatures. Magenta arrow indicates the energy gap at 80 K.}
		\label{fig5}
	\end{figure}

	$Appendix\,\,-$  
	While dispersive spin waves of YbFe$_6$Ge$_6$ extend to much higher energies, only the low-energy part (i.e., the spin-wave gap) shows a significant change across $T_{\rm{SR}}$. Fig. \ref{fig5}(a) and (b) show the excitation spectra along the [0, 0, $L$] direction at 10 K and 100 K, respectively \cite{SM}. The signals around $L$ = 1 and 3 are magnons, while phonons appear at larger momentum transfer positions. The overall excitation spectrum changes little except for the intensity, as can be seen from the constant-energy cuts shown in Fig. \ref{fig5}(c) and (d). We notice the intensity increase from 10 K to 100 K is significantly larger than the effect simply due to the Bose factor [grey curves in Fig. \ref{fig5}(c) and (d)], indicating a change of the spin orientation. Thermal evolution of the low-energy spin excitations around (0, 0, 1) is presented in Fig. \ref{fig5}(e)-(i) \cite{SM}. Since the magnon has a bandwidth of $\sim$40 meV along the [0, 0, $L$] direction \cite{SM}, the spin excitations below 3 meV are rod-like and concentrated at the Brillouin zone center. As the temperature rises to near $T_{\rm{SR}}$, their intensity gradually accumulates, which extends down to the elastic background [Fig. \ref{fig5}(e)-(h)]. However, the intensity around 1 meV drops at 80 K ($> T_{\rm{SR}}$) [Fig. \ref{fig5}(i)], indicating the presence of an energy gap, consistent with the data shown in Fig. \ref{fig4}.
	
	%%%%%%%%%% Merge with supplemental materials %%%%%%%%%%
	\pagebreak
	\pagebreak
	
	\widetext
	
	\begin{center}
		\textbf{\large Supplemental Material for ``Anomalous Electrical Transport in the Kagome Magnet \ch{YbFe_6Ge_6}''}
	\end{center}

	%%%%%%%%%% Merge with supplemental materials %%%%%%%%%%
	%%%%%%%%%% Prefix a "S" to all equations, figures, tables and reset the counter %%%%%%%%%%
	\setcounter{equation}{0}
	\setcounter{table}{0}
	\setcounter{page}{1}
	\makeatletter
	\renewcommand{\theequation}{S\arabic{equation}}
	\renewcommand{\thefigure}{S\arabic{figure}}
	%\renewcommand{\bibnumfmt}[1]{[S#1]}
	%\renewcommand{\citenumfont}[1]{S#1}
	%%%%%%%%%% Prefix a "S" to all equations, figures, tables and reset the counter %%%%%%%%%%
	
	\section{Single crystal growth and characterization}
\begin{figure}[b!]
		\centering{\includegraphics[clip,width=9cm]{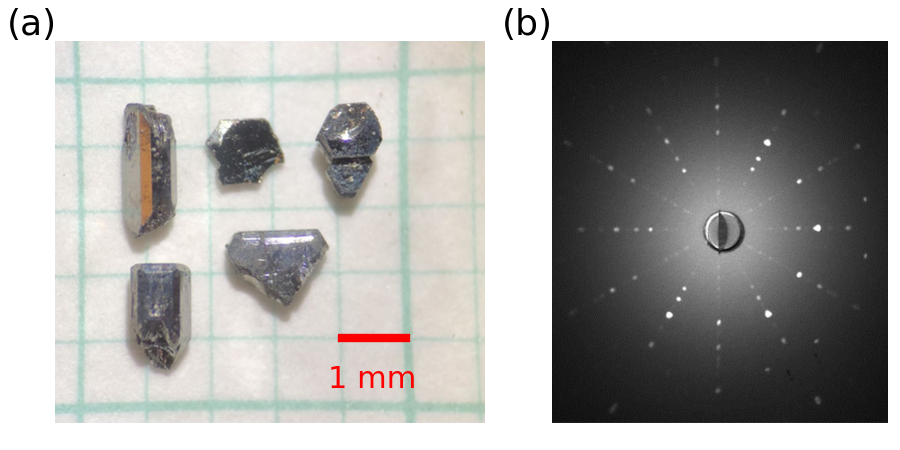}}
		\caption{(a) Representative \ch{YbFe_6Ge_6} single crystals against a millimeter grid. (b) X-ray Laue diffraction pattern taken on the (0, 0, $L$) surface.}
		\label{figs1}
	\end{figure}

\begin{figure}[t!]
		\centering{\includegraphics[clip,width=9cm]{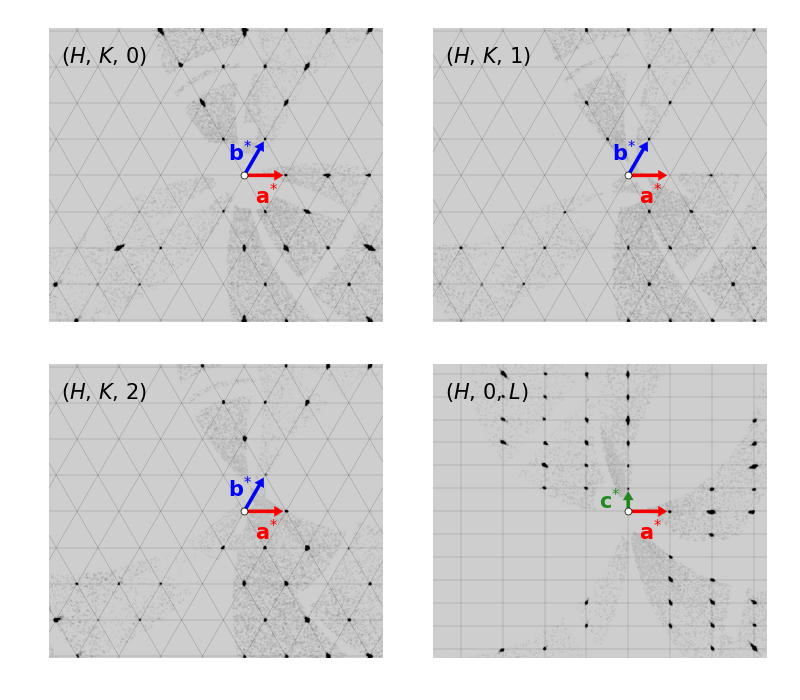}}
		\caption{Single crystal XRD patterns of the ($H$, $K$, 0), ($H$, $K$, 1), ($H$, $K$, 2), and ($H$, 0, $L$) planes. Arrows represent primitive vectors ($\textbf{a}^{*}$, $\textbf{b}^{*}$, and $\textbf{c}^{*}$) of the reciprocal space.}
		\label{figs2}
	\end{figure}

\begin{figure}[h!]
		\centering{\includegraphics[clip,width=7cm]{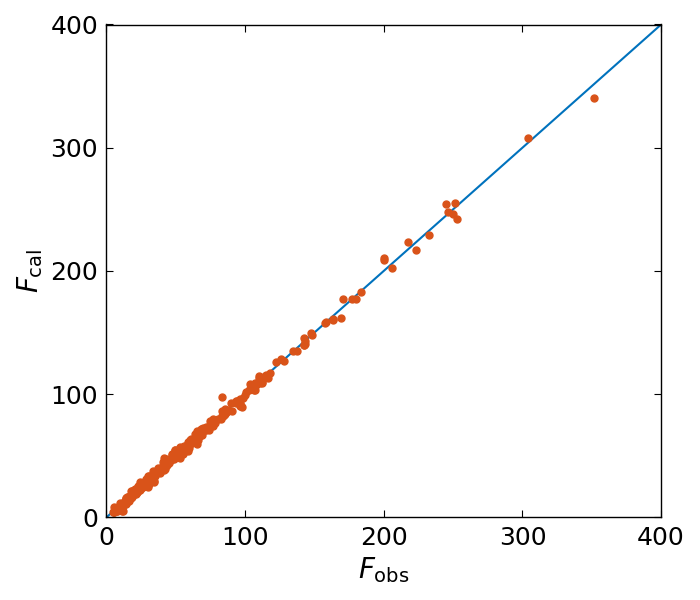}}
		\caption{Calculated structure factor F$_{\rm{cal}}$ versus observed structure factor F$_{\rm{obs}}$ from single crystal XRD measured at 100 K.}
		\label{figs3}
	\end{figure}

\begin{figure}[t!]
		\centering{\includegraphics[clip,width=9cm]{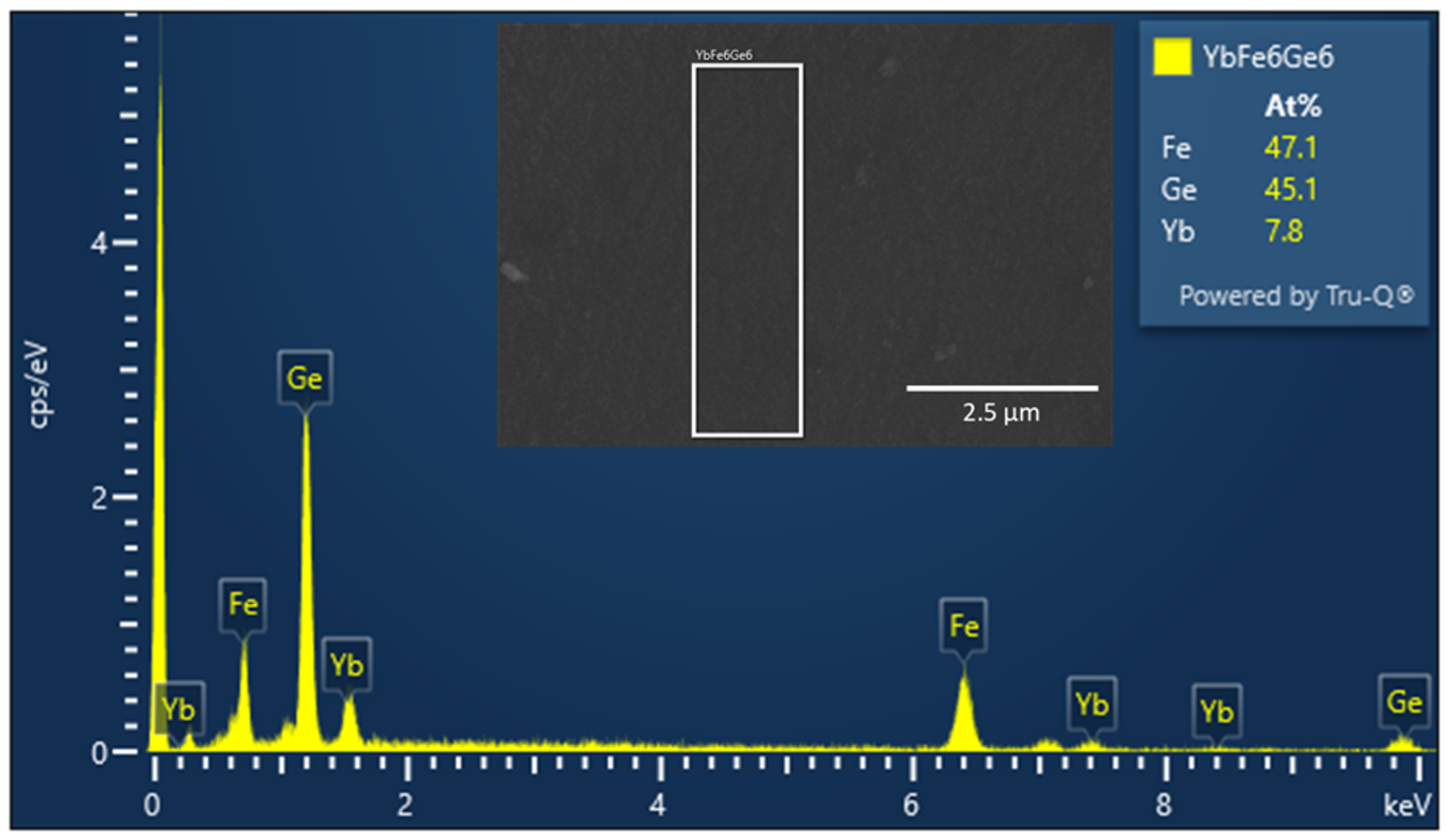}}
		\caption{Energy dispersive X-ray spectrum at a representative spot on a \ch{YbFe_6Ge_6} single crystal. Inset shows the image taken by scanning electron microscopy.}
		\label{figs4}
	\end{figure}

	\ch{YbFe_6Ge_6} single crystals were grown with a Sn-flux method \cite{AvilaJPCM2005}. Raw materials of Yb, Fe, Ge, and Sn powders with a molar ratio of 1 : 1 : 3 : 20 were mixed and loaded into a quartz tube in a glove box. A piece of quartz wool was put in the upper part of the quartz tube before evacuating and sealing. The sealed quartz tube was put into a box furnace and heated to 1100 $^{\circ}$C in 24 hours. After keeping at 1100 $^{\circ}$C for 10 hours, the quartz tube was gradually cooled at a rate of 5 $^{\circ}$C/hour until 500 $^{\circ}$C, at which it was flipped and centrifuged to remove excess Sn. \ch{YbFe_6Ge_6} single crystals in hexagonal flake or prism can be obtained on the quartz wool, with a typical size of 1$\sim$2 millimeters [see Fig. \ref{figs1}(a)]. The crystals were soaked in HCl solution for several days to remove residual Sn. Sharp and clear diffraction spots in the X-ray Laue (Photonic Science) diffraction pattern confirm the high quality of these crystals Fig. \ref{figs1}(b). To check the crystal structure, single-crystal X-ray diffraction (XRD) was conducted with the Rigaku Synergy-S diffractometer. Fig. \ref{figs2} shows representative XRD patterns. Crystal structure refinement was performed with the Jana2006 program \cite{Petricek2014}. The obtained space group is $P$6/$mmm$. The comparison between the observed and calculated structure factors is presented in Fig. \ref{figs3}. The crystallographic information can be found in Table \ref{tb1} and Table \ref{tb2}. The chemical elements of our sample were characterized by energy dispersive X-ray spectrum (EDS) in an EMC Helios 660 scanning electron microscope (SEM) (Fig. \ref{figs4}). The elemental analysis shown in Table \ref{tb3} indicates the chemical elements of our sample are in good agreement with the chemical formula \ch{YbFe_6Ge_6}.
	
	\section{Magnetization measurements}
	
	Magnetization measurements up to 9 T were taken with a Quantum Design PPMS system. Pulsed high-field magnetization measurements up to 50 T were performed at Institute for Solid State Physics (ISSP), the University of Tokyo. The absolute magnetizations were normalized according to the magnetization data taken with the Quantum Design PPMS system.
	
	\begin{figure}[b!]
		\centering{\includegraphics[clip,width=16cm]{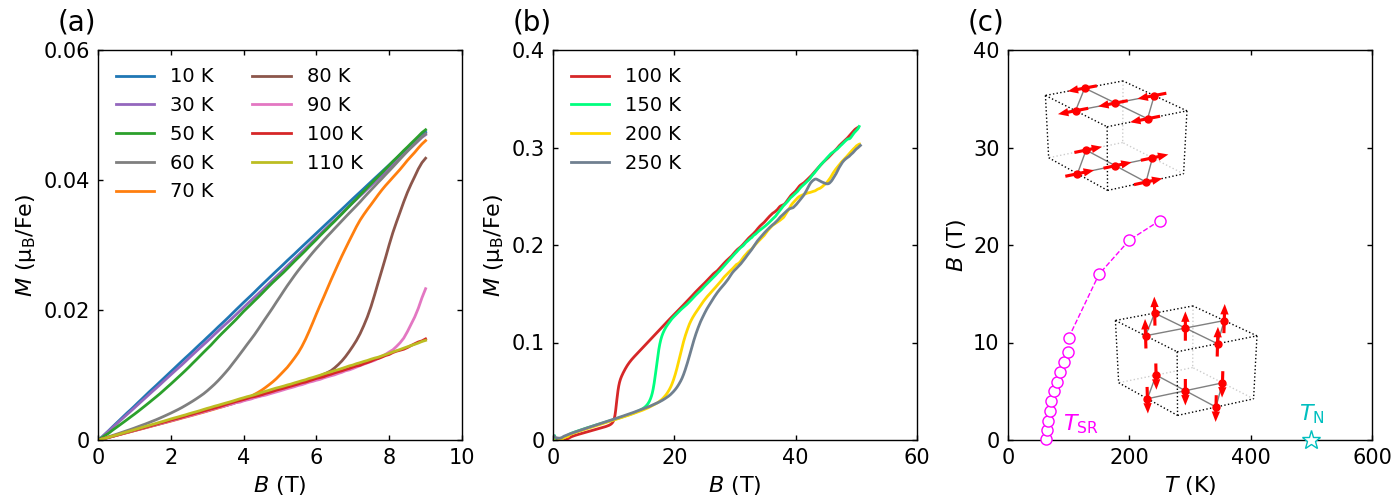}}
		\caption{(a) Isothermal magnetizations along the $c$-axis up to 9 T at selected temperatures. (b) Pulsed high-field magnetizations along the $c$-axis up to 50 T at selected temperatures. The oscillation features at high magnetic fields are due to sample vibration at high temperatures (200 K and 250 K). (c) Phase diagram about $c$-axis magnetic field and temperature. Inset shows the magnetic structures on both sides of the phase boundary.}
		\label{figs5}
	\end{figure}
	
	\begin{figure}[t!]
		\centering{\includegraphics[clip,width=9cm]{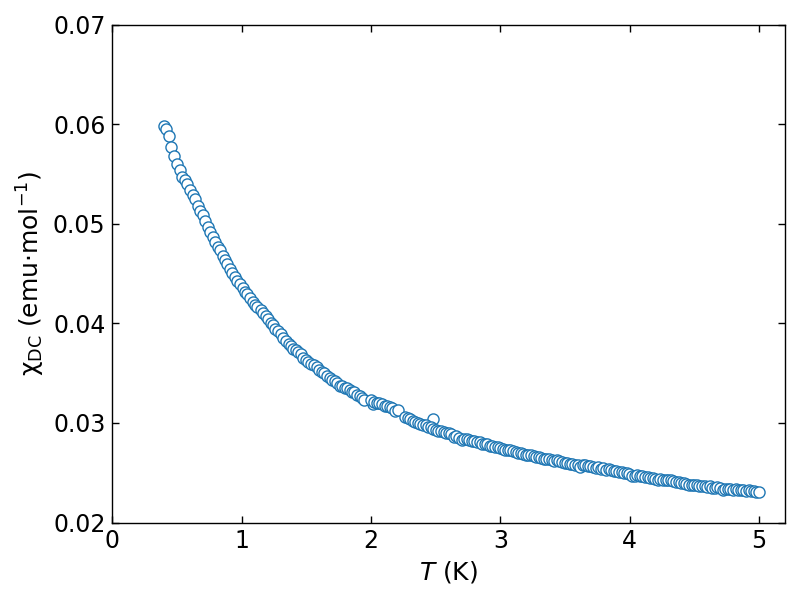}}
		\caption{Temperature dependence of DC magnetic susceptibility from 0.4 K to 5 K under a field of 0.1 T applied parallel to the $c$-axis.}
		\label{figs6}
	\end{figure}
	
	Fig. \ref{figs5}(a) presents the isothermal magnetizations along the $c$-axis. At low temperatures below $T_{\rm{SR}}$, the magnetization is almost linear with respect to the field. However, when the temperature goes above $T_{\rm{SR}}$, the magnetization shows a significant slope change, which is due to the field-induced spin reorientation (SR, or spin flop) transition. The transition field gradually increases as the temperature goes up and is larger than 9 T when the temperature is above 100 K, which can be confirmed by the pulsed high-field magnetizations shown in Fig. \ref{figs5}(b). By collecting the transition temperatures at fixed fields and transition fields at fixed temperatures, we construct the phase boundary about $c$-axis magnetic field and temperature [Fig. \ref{figs5}(c)]. Due to the spontaneous SR transition, the phase boundary ends at the temperature axis. This phase diagram is consistent with the previous report \cite{AvilaJPCM2005}. Note that $T_{\rm{SR}}$ does not depend on the magnetic fields applied along the kagome plane (i.e., the phase boundary is parallel to the field axis) \cite{AvilaJPCM2005}.
	
	To check whether the Yb spins order at lower temperatures, additional DC magnetic susceptibility was measured using a Quantum Design MPMS with the He3 option. From Fig. \ref{figs6}, we can see the magnetic susceptibility shows a smooth upturning on cooling from 5 K to 0.4 K, which is typical of paramagnetic behavior. Therefore, the Yb spins in this system remain disordered down to 0.4 K, indicating rather weak Yb-Yb interaction. For the magnetic and electrical transport properties above 2 K, which are the focus of the present study, the interaction between Yb and Fe is most relevant.

	\section{Electrical transport measurements}
	
	Electrical transport measurements up to 9 T were performed in a Quantum Design PPMS system. Electrical transport measurements up to 12 T were performed in a cryogen-free superconducting magnet (TeslatronPT, Oxford) with a homemade stick. YbFe$_6$Ge$_6$ single crystals were cut and polished into an approximately rectangular shape before making electrical contacts. The typical sample size in our transport measurement is about 1 $\times$ 0.5 $\times$ 0.15 mm$^3$. Contact misalignment was corrected by field symmetrizing and anti-symmetrizing the longitudinal and transverse resistance data, respectively.
	
	\subsection{Electrical transport under in-plane magnetic fields}
	
	\begin{figure}[b!]
		\centering{\includegraphics[clip,width=15cm]{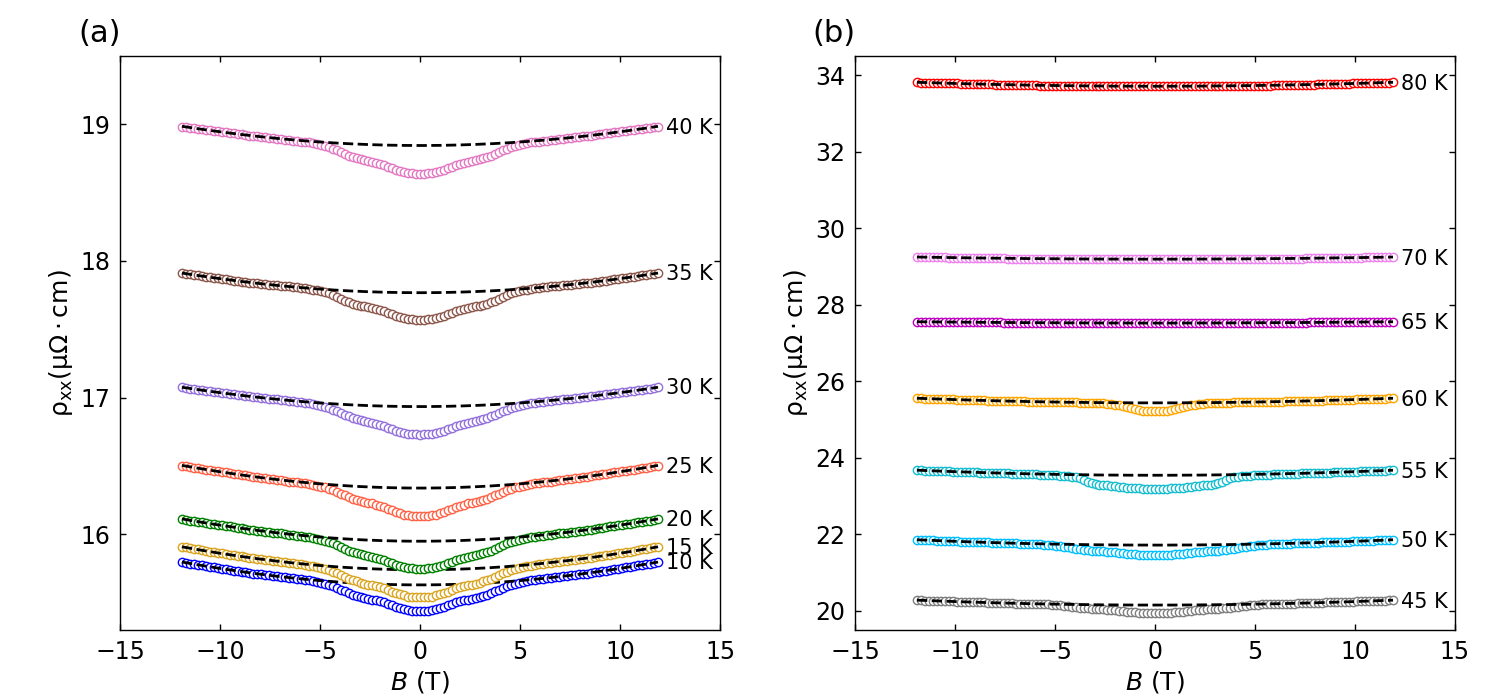}}
		\caption{Field dependence of longitudinal resistivity at selected temperatures with the field applied along the \textbf{a}$^\ast$ direction. The electric current is along the $b$-axis. Dashed curves are the fits with the two-band model.}
		\label{figs7}
	\end{figure}
	
	\begin{figure}[b!]
		\centering{\includegraphics[clip,width=14cm]{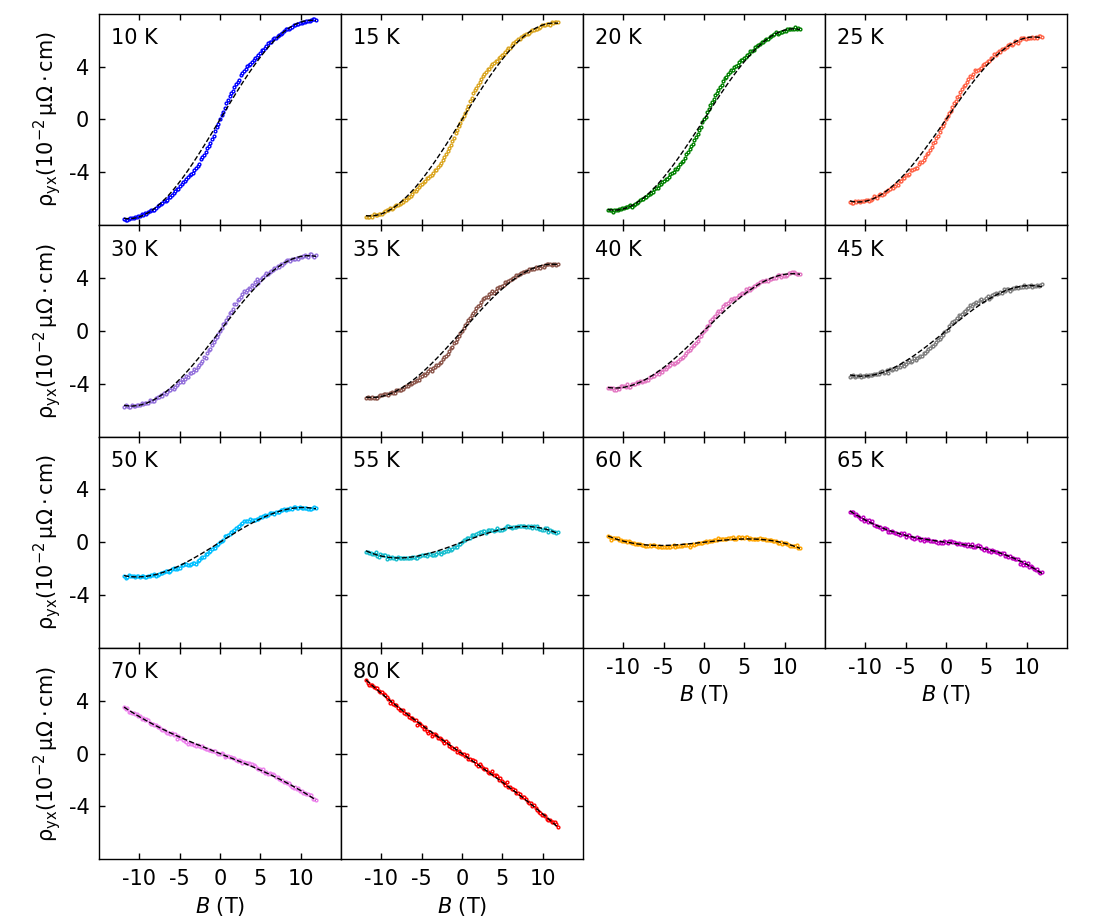}}
		\caption{Hall resistivity at selected temperatures with the field applied along the \textbf{a}$^\ast$ direction. Dashed curves are the fits with the two-band model.}
		\label{figs8}
	\end{figure}
	
\begin{figure}[h!]
		\centering{\includegraphics[clip,width=15cm]{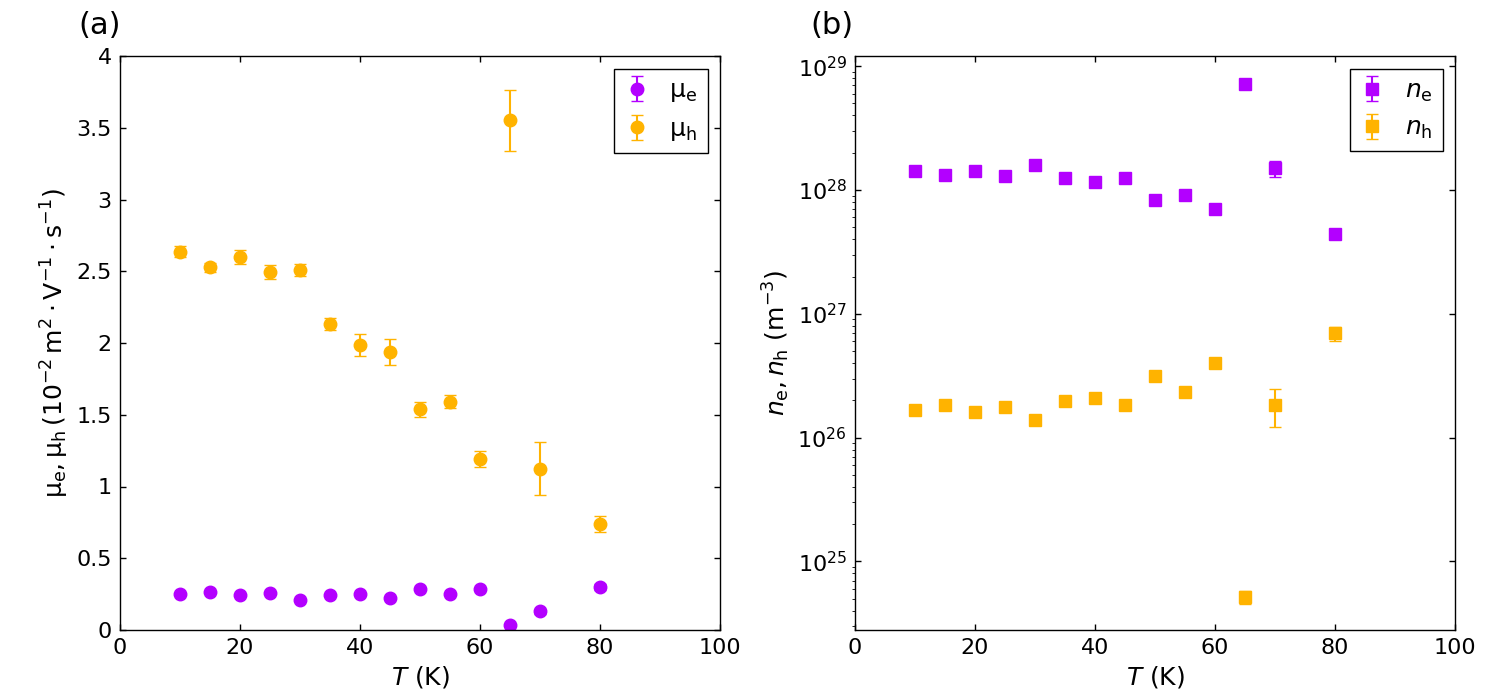}}
		\caption{Temperature dependence of the mobility (a) and carrier density (b) obtained from the two-band model fit.}
		\label{figs9}
	\end{figure}

\begin{figure}[t!]
		\centering{\includegraphics[clip,width=7cm]{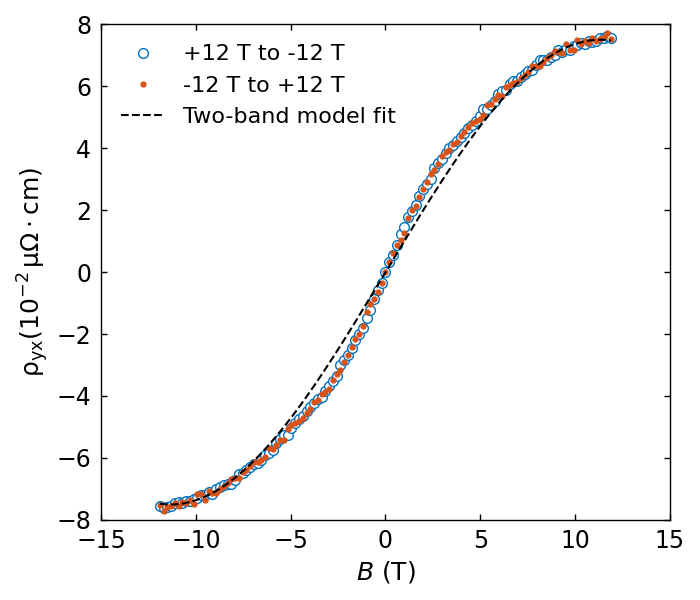}}
		\caption{Hall resistivity at 10 K measured from +12 T to -12 T (blue) and from -12 T to +12 T (orange). The dashed curve is the fit with the two-band model.}
		\label{figs10}
	\end{figure}

	Magnetoresistance (MR) and Hall measurements at 14 temperatures were shown in Fig. \ref{figs7} and Fig. \ref{figs8}, which include the data shown in Fig. 2(a) and (b) of the main text. The magnetic fields were applied along the $\mathbf{a}^{\ast}$ direction. We use a two-band model to simultaneously fit the MR and Hall data above 6 T \cite{Ashcroft1976,WangPRL2024,ZhuJPCM2018}, according to which the longitudinal and transverse resistivities can be written as
	\begin{equation}
		\rho_{\rm{xx}}^{\rm{O}}(B) = \frac{1}{e} \frac{(n_{\rm{h}}\mu_{\rm{h}}+n_{\rm{e}}\mu_{\rm{e}})+(n_{\rm{h}}\mu_{\rm{e}}+n_{\rm{e}}\mu_{\rm{h}})\mu_{\rm{h}}\mu_{\rm{e}}B^2}{(n_{\rm{h}}\mu_{\rm{h}}+n_{\rm{e}}\mu_{\rm{e}})^2+(n_{\rm{h}}-n_{\rm{e}})^2\mu_{\rm{h}}^2\mu_{\rm{e}}^2B^2}
	\end{equation}
	and
	\begin{equation}
		\rho_{\rm{yx}}^{\rm{O}}(B) = \frac{1}{e} \frac{(n_{\rm{h}}\mu_{\rm{h}}^2-n_{\rm{e}}\mu_{\rm{e}}^2)+(n_{\rm{h}}-n_{\rm{e}})\mu_{\rm{h}}^2\mu_{\rm{e}}^2B^2}{(n_{\rm{h}}\mu_{\rm{h}}+n_{\rm{e}}\mu_{\rm{e}})^2+(n_{\rm{h}}-n_{\rm{e}})^2\mu_{\rm{h}}^2\mu_{\rm{e}}^2B^2}B,
	\end{equation}
	where $e$ represents the elementary charge, $n_{\rm{h}}$ and $n_{\rm{e}}$ are the carrier densities of the hole-like and electron-like bands, respectively, and $\mu_{\rm{h}}$ and $\mu_{\rm{e}}$ are the corresponding mobilities. The fitted results are shown as the dashed curves in Fig. \ref{figs7} and Fig. \ref{figs8}. The temperature dependences of the mobility and carrier density are presented in Fig. \ref{figs9}. Below $T_{\rm{SR}}$, the hole-like band has a larger mobility than the electron-like band, but its carrier density is much smaller than the latter. Such behavior is not uncommon in metals with magnetic rare earth ions \cite{YePRB2017,RamPRB2023}. As the temperature increases, the difference between the two kinds of bands narrows, indicating the weakened impact of the rare earth ion (Yb$^{3+}$) \cite{AvilaJPCM2005} and the material becomes closer to a one-band-like system at high temperature. An anomaly appears around $T_{\rm{SR}}$, which is consistent with the SR transition and the derivative of resistivity [Fig. 1(b) of the main text]. Here we need to point out that the actual electronic structure of \ch{YbFe_6Ge_6} might be much more complex than a two-band system. For these electrical transport measurements, no hysteresis can be discerned between field ramp-up and ramp-down. For example, the Hall resistivities measured from +12 T to -12 T and -12 T to +12 T basically overlap with each other [see Fig. \ref{figs10}]. The anomalous Hall resistivity is obtained by subtracting the fitted curve of the two-band model (ordinary Hall resistivity $\rho_{\rm{yx}}^{\rm{O}}$)
	\begin{equation}
		\Delta \rho_{\rm{yx}} = \rho_{\rm{yx}} - \rho_{\rm{yx}}^{\rm{O}}.
	\end{equation}
	The anomalous Hall conductivity presented in Fig. 2(c) and (d) of the main text is approximately calculated as
	\begin{equation}
		\Delta \sigma_{\rm{xy}} = \frac{\Delta \rho_{\rm{yx}}}{\rho_{\rm{xx}}^2 + \rho_{\rm{yx}}^2}.
	\end{equation}
	
\begin{figure}[t!]
		\centering{\includegraphics[clip,width=11cm]{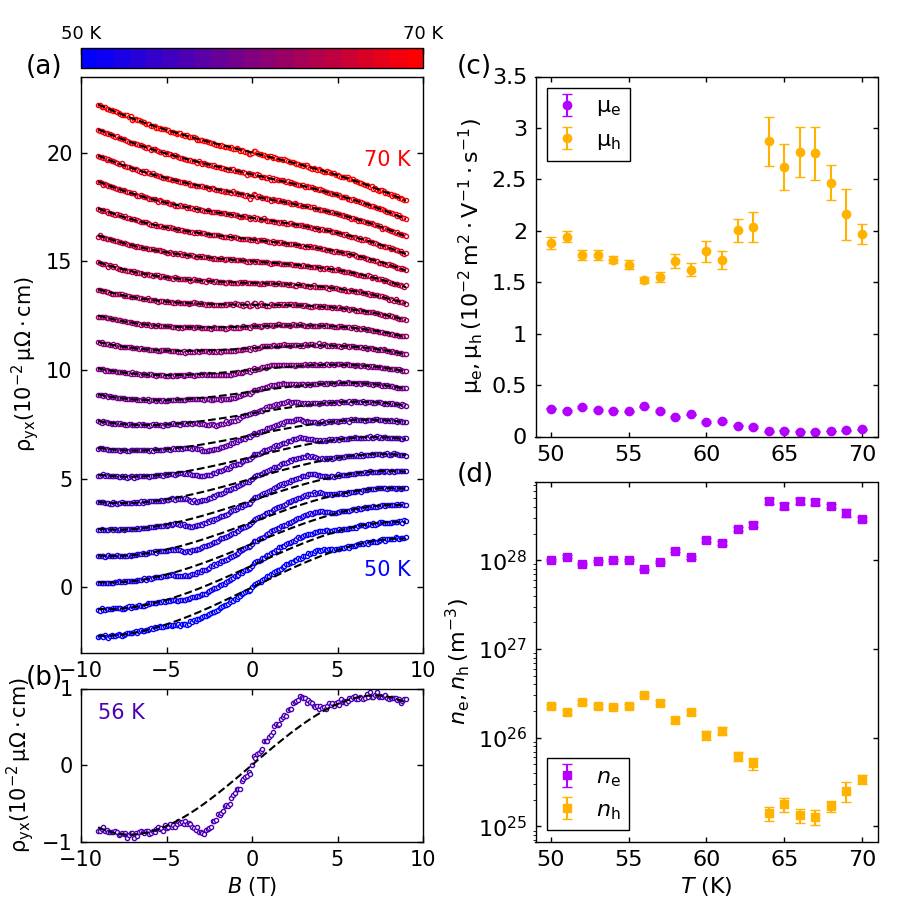}}
		\caption{(a) Hall resistivity from 50 K to 70 K with 1 K per step. Magnetic field up to $\pm$9 T is applied along the \textbf{a}$^\ast$ direction. The electric current is along the $b$-axis. Dashed curves are the fits with the two-band model. Data above 50 K are offset by 0.01 $\mu \Omega \cdot \rm{cm}$ for clarity. (b) Same data at 56 K in (a). (c) and (d) Temperature dependence of the mobility and carrier density of the two-band model fit.}
		\label{figs11}
	\end{figure}

	To further confirm the observed anomalous Hall effect (AHE), additional electrical transport measurements were conducted with the Quantum Design PPMS system. Fig. \ref{figs11}(a) presents detailed Hall resistivities from 50 K to 70 K. In this temperature region, the curved multi-band feature is evident even for fields below 9 T, which enables us to fit the data with a two-band model. Below $T_{\rm{SR}}$, we can clearly see that the Hall resistivity at low field deviates from the curve of the two-band model. This deviation, however, gradually smears out as the temperature goes up. Fig. \ref{figs11}(b) shows an enlarged view of the data and the fit at 56 K. The resultant fitting parameters are shown in Fig. \ref{figs11}(c) and (d), which are reasonably consistent with the data measured up to 12 T (Fig. \ref{figs9}).
	
\begin{figure}[h!]
		\centering{\includegraphics[clip,width=13cm]{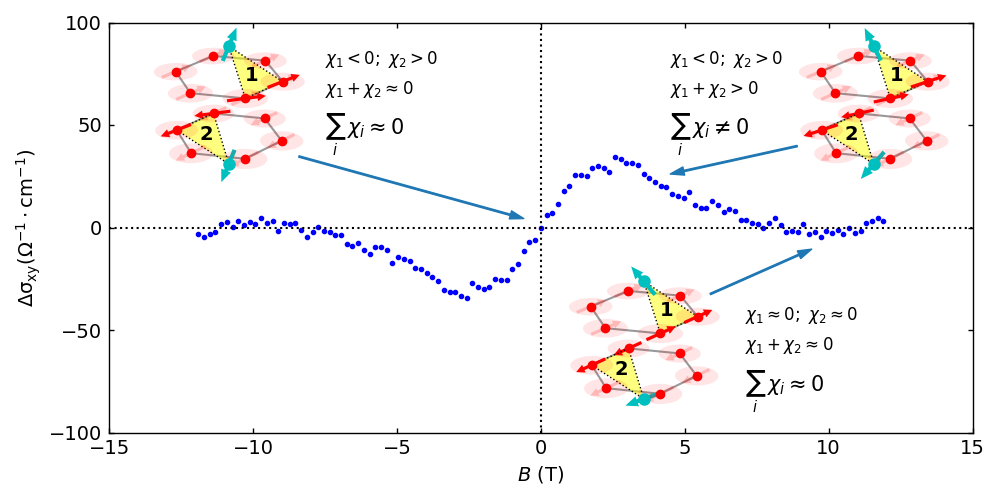}}
		\caption{Anomalous Hall conductivity $\Delta\sigma_{xy}$ at 10 K. Insets schematically show the behaviors of the spin configurations of Yb (cyan) and Fe (red), respectively, at various magnetic fields, as indicated by the blue arrows. Yellow triangles (``1'' and ``2'') highlight two spin triads of Yb-Fe. Light red arrows represent the static magnetic structure below $T_{\rm{SR}}$. Light red circles indicate the ``easy plane'' for the Fe spins. Spin canting angles are exaggerated for better visualization.}
		\label{figs12}
	\end{figure}

	The observed AHE below $T_{\rm{SR}}$ can be understood through a dynamic scalar spin chirality (SSC) scattering mechanism, which is schematically illustrated by Fig. \ref{figs12}. Due to the existence of the Yb spins and their increased interaction with the Fe kagome lattice below $T_{\rm{SR}}$, two adjacent Fe spins can form a spin triad with the out-of-plane Yb spin (yellow triangles in Fig. \ref{figs12}). When considering the fluctuations of the Fe spins from a semiclassical perspective, they will deviate from the static collinear state. So this kind of Yb-Fe spin triad can have non-zero SSC. Without an external magnetic field, these local SSCs will cancel out due to the combined space inversion and time-reversal ($\mathcal{IT}$) symmetry. For example, in Fig. \ref{figs12}, spin triad 1 has negative SSC $\chi_1 < 0$, but spin triad 2, connected by the global $\mathcal{IT}$ symmetry, has a positive SSC $\chi_2 > 0$ with the same magnitude. These local SSCs therefore do not contribute an overall SSC, i.e., $\sum_{i}\chi_i \approx 0$, which results in the absence of AHE at zero field.
	
	Applying a magnetic field slightly aligns the Yb spins to the field direction, which breaks the balance between positive and negative SSCs, despite low-energy spin fluctuations being slightly suppressed by the field. The static Fe spin orientation is almost not affected by the field, as the energy scale of Fe-Fe interaction is much larger than the magnetic field. This produces a net overall SSC ($\sum_{i}\chi_i \neq 0$) so that gives rise to an AHE. However, when the magnetic field becomes very large, low-energy spin fluctuations are significantly suppressed by opening a Zeeman gap. The spin fluctuation induced SSC, along with the accompanied AHE, then vanishes again. This process qualitatively explains the field dependence of the observed AHE. Here, the change of spin anisotropy in YbFe$_6$Ge$_6$ is critical, where the Yb-Fe interaction at low temperatures triggers the change of spin anisotropy and gives rise to the gapless spin excitations. Spin anisotropy change is also found to be important for the spin-fluctuation driven AHE in the ferromagnetic heterostructure SrRuO$_3$/SrTiO$_3$ \cite{WangNM2019}. However, in that case, it is due to the thermal fluctuation near the ferromagnetic transition temperature, which is different from our case.
	
	\subsection{Electrical transport under out-of-plane magnetic fields}
	
\begin{figure}[b!]
		\centering{\includegraphics[clip,width=10cm]{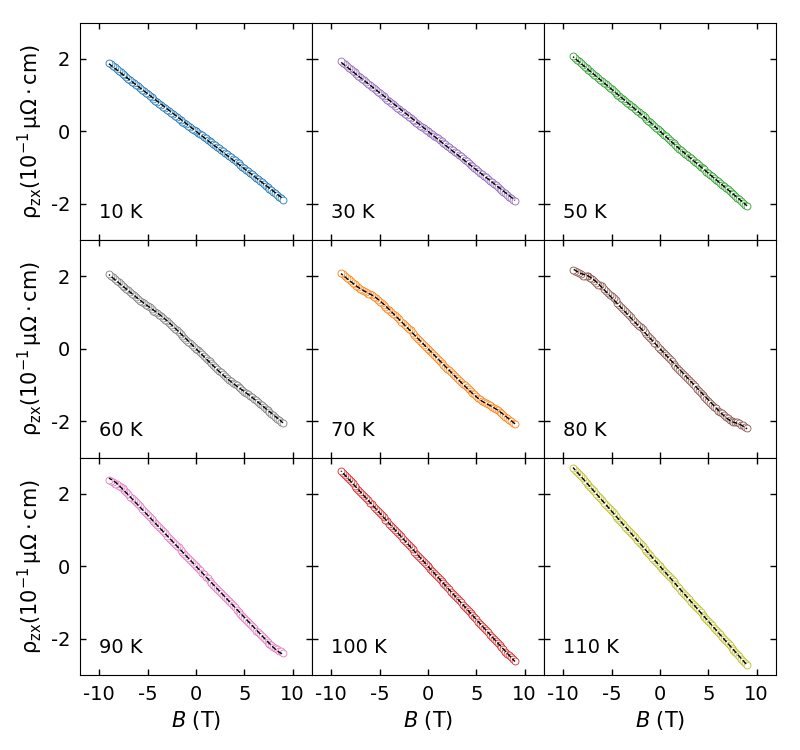}}
		\caption{Hall resistivity at selected temperatures with the field applied along the $c$-axis. Dashed curves are the fits as described in the text.}
		\label{figs13}
	\end{figure}

\begin{figure}[b!]
		\centering{\includegraphics[clip,width=10cm]{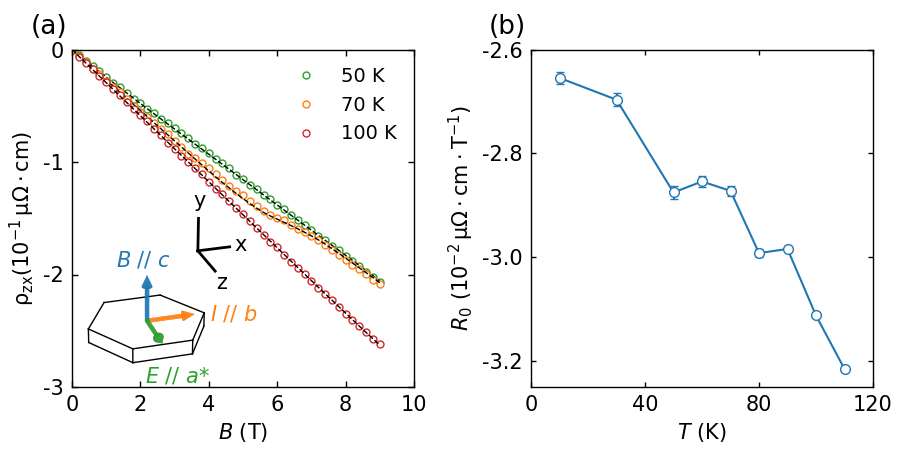}}
		\caption{Hall resistivity at 50 K, 70 K, and 100 K with the field applied along the $c$-axis. Dashed curves are the fits as described in the text. Inset shows the Hall measurement configuration and the definition of $x$, $y$, and $z$ axes. (b) Temperature dependence of the Hall coefficient $R_0$ obtained from the fitting.}
		\label{figs14}
	\end{figure}

	Fig. \ref{figs13} shows the Hall resistivity $\rho_{\rm{zx}}$ at selected temperatures from 10 K to 100 K under $c$-axis magnetic fields. Although the Hall resistivity is largely linear with respect to the field, it shows slope change for temperatures above $T_{\rm{SR}}$, as can be more clearly seen in Fig. \ref{figs14}(a). This is similar to the magnetization data of this field direction [Fig. \ref{figs5}(a) in Section II]. Note that for high temperatures, the slope change of $\rho_{\rm{zx}}$ is expected to occur at magnetic fields larger than 9 T, which is beyond our measurements. Such behavior indicates a non-negligible contribution to the Hall effect from the net magnetization along the $c$-axis.
	
	We find the measured $\rho_{{zx}}$ can be well fitted with the following equation
	\begin{equation}
		\rho_{{zx}} = \rho^{\rm{O}}_{{zx}} + \rho^{\rm{A}}_{{zx}},
	\end{equation}
	where $\rho^{\rm{O}}_{{zx}}$ is ordinary Hall resistivity and $\rho^{\rm{A}}_{{zx}}$ is conventional anomalous Hall resistivity due to net magnetization. They can be written as
	\begin{equation}
		\rho^{\rm{O}}_{{zx}} = R_0B
	\end{equation}
	and
	\begin{equation}
		\rho^{\rm{A}}_{{zx}} = S_{\rm{A}}M.
	\end{equation}
	$R_0$ is the ordinary Hall coefficient and $S_{\rm{A}}$ is another proportional coefficient. Since both the Hall resistivity and magnetization are almost proportional to the magnetic field at low and high temperatures (i.e., 10 K, 30 K, 50 K, 100 K, and 110 K), we cannot determine $R_0$ and $S_{\rm{A}}$ separately. To add constraints, we assume $S_{\rm{A}}$ is temperature independent \cite{NagaosaRMP2010} and simultaneously fit the data at 9 temperatures. The fitted results are shown as the dashed curves in Fig. \ref{figs13} and Fig. \ref{figs14}(a). The ordinary Hall coefficient $R_0$ is shown in Fig. \ref{figs14}(b) and $S_{\rm{A}}$ is 1.12(2) $\mu \Omega \cdot \rm{cm} \cdot \mu_{\rm{B}}^{-1} \cdot \rm{Fe}$. Therefore, no additional spin-fluctuation driven AHE contributes to the $\rho_{{zx}}$, where the magnetic field is applied along the $c$-axis. This observation is similar to other kagome magnets with easy-plane magnetic structures - $A$Mn$_6$Sn$_6$ ($A$ = Y, Sc, and Er) \cite{GhimireSA2020,ZhangAPL2022,FruhlingPRM2024} and HoAgGe \cite{RoychowdhuryPNAS2024}, where the spin-fluctuation driven AHEs can only be observed when the magnetic fields are applied along the kagome plane. It suggests that these AHEs are primarily contributed by in-plane spin fluctuations \cite{GhimireSA2020,ZhangAPL2022,FruhlingPRM2024,RoychowdhuryPNAS2024}, which are suppressed by an out-of-plane field as the spins tilt away from the magnetic easy plane.
	
	\section{Neutron diffraction and analysis of the magnetic structure}
	
	\subsection{Neutron diffraction measurements}
	Neutron diffraction measurements were conducted with the CORELLI diffractometer at Spallation Neutron Source, Oak Ridge National Laboratory \cite{YeJAC2018} and the DMC diffractometer at Swiss Spallation Neutron Source (SINQ), Paul Scherrer Institut. One piece of YbFe$_6$Ge$_6$ single crystal with a mass of about 15 mg was used in these two experiments, respectively.
	
\begin{figure}[h!]
		\centering{\includegraphics[clip,width=11cm]{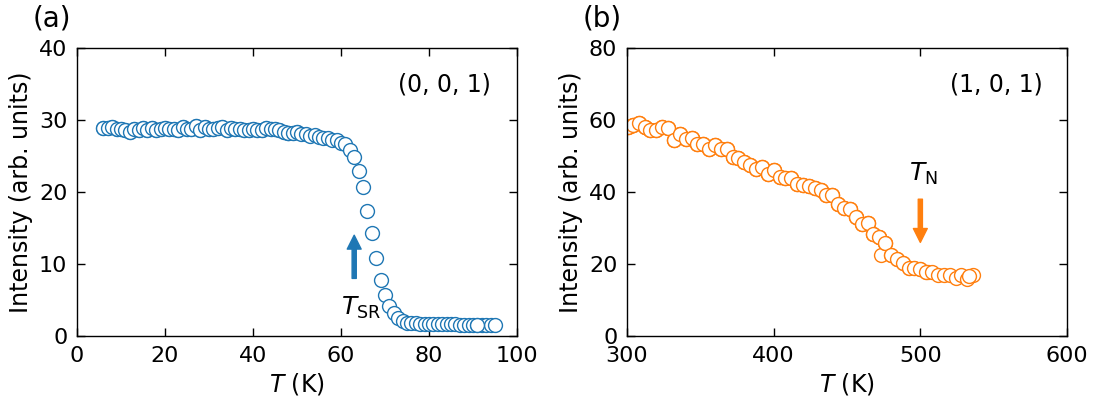}}
		\caption{Temperature dependence of the diffraction intensities for Bragg peaks (0, 0, 1) (a) and (1, 0, 1) (b), where arrows indicate $T_{\rm{SR}}$ and $T_{\rm{N}}$, respectively.}
		\label{figs15}
	\end{figure}

\begin{figure}[h!]
		\centering{\includegraphics[clip,width=11cm]{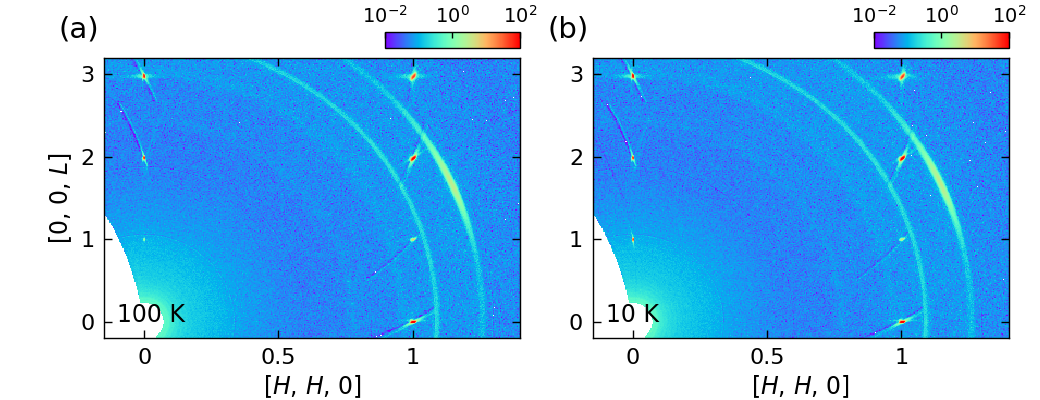}}
		\caption{Neutron diffraction patterns for the $(H, H, L)$ plane taken on a piece of YbFe$_6$Ge$_6$ single crystal at 100 K (a) and 10 K (b), respectively. The intensity color bar is in log scale to highlight possible weak features.}
		\label{figs16}
	\end{figure}

	Fig. \ref{figs15}(a) shows the temperature dependence of the diffraction intensity at (0, 0, 1). We can see the intensity abruptly drops when the temperature goes above $T_{\rm{SR}} \approx$ 63 K, which is due to the disappearance of the spin component along the $c$-axis. At the Bragg peak (1, 0, 1), the intensity shows a canonical order-parameter behavior with respect to the temperature [Fig. \ref{figs15}(b)], consistent with the ordering temperature $T_{\rm{N}} \approx$ 500 K. These temperature-dependent data were taken at the CORELLI diffractometer. The absence of additional Bragg peaks below $T_{\rm{SR}}$ can be further confirmed by the diffraction pattern taken at ($H, H, L$) plane [Fig. \ref{figs16}], which corroborates the observation that neither charge order nor incommensurate magnetic order occurs below $T_{\rm{SR}}$. Diffraction maps presented in Fig. 3 of the main text and Fig. \ref{figs16} were taken with the DMC diffractometer.
	
	\subsection{Analysis of the magnetic structure}
	
	With the magnetic propagation wave vector $\textbf{k}_{\rm{m}}$ = (0, 0, 0) in hand, we can do representation analysis to get possible magnetic structures \cite{Wills2001}. For this purpose, we use the BasIReps program implemented in the FullProf suite \cite{Rodriguez-Carvajal1993}. The magnetic representation $\Gamma_{\rm{mag}}$ of the Fe ion located at (0.5, 0, $z_{\rm{Fe}}$) (with $z_{\rm{Fe}}$ = 0.2491) for space group $P$6/mmm can be decomposed as
	\begin{equation}
		\Gamma_{\rm{mag}}=1\Gamma_{2}^1+1\Gamma_{3}^1+1\Gamma_{4}^1+1\Gamma_{5}^2+2\Gamma_{6}^2+1\Gamma_{7}^1+1\Gamma_{9}^1+1\Gamma_{10}^1+1\Gamma_{11}^2+2\Gamma_{12}^2.
	\end{equation}
	The irreducible representations (IRs) and their basis vectors (BVs) are listed in Table \ref{tb4}. Assuming Fe ions have the same magnetic moment size, we can construct all possible magnetic structures from these BVs, which are explicitly shown in Fig. \ref{figs11}. Among them, ferromagnetic structures $\Psi_{1}$ [Fig. \ref{figs11}(a)] and $\Psi_{8}-\Psi_{9}$ [Fig. \ref{figs11}(e)] are irrelevant to the magnetic structure of YbFe$_6$Ge$_6$, which is an antiferromagnet. $\Psi_{10}$ [Fig. \ref{figs11}(f)] corresponds to the $A$-type antiferromagnetic structure between $T_{\rm{N}}$ and $T_{\rm{SR}}$. For the remaining 7 configurations, their magnetic structure factors at Bragg peak positions can be directly calculated and compared with the (integrated) intensities obtained from neutron diffraction.
	
	In neutron diffraction, the integrated intensity of a magnetic Bragg peak is proportional to the modulus square of the magnetic structure factor $\textbf{F}_{\rm{M}}(\textbf{Q})$ \cite{Shirane2002}
	\begin{equation}
		I_{\rm{M}}(\textbf{Q})=A\left.\frac{\rm{d\sigma_{\rm{M}}}}{\rm{d\Omega}}\right|_{\textbf{Q}}=AN_{\rm{M}}\frac{{(2\pi)}^3}{V_{\rm{M}}}|\textbf{F}_{\rm{M}}(\textbf{Q})|^2,
	\end{equation}
	where $N_{\rm{M}}$ is the number of magnetic unit cells in the sample, $V_{\rm{M}}$ is the volume of the magnetic unit cell. The magnetic structure factor can be written as
	\begin{equation}
		\textbf{F}_{\rm{M}}(\textbf{Q})=\sum_{j}\frac{\gamma r_0}{2} g_jf_j(Q)\textbf{S}_{\perp j}e^{i\textbf{Q}\cdot\textbf{r}_j}e^{-W_j}.
	\end{equation}
	$\textbf{S}_{\perp j}$ is the spin size at site $j$ that is detectable by neutrons
	\begin{equation}
		\textbf{S}_{\perp j}=\textbf{S}_j-\hat{\textbf{Q}}(\hat{\textbf{Q}}\cdot\textbf{S}_j),
	\end{equation}
	where $\hat{\textbf{Q}}$ is the unit vector of $\textbf{Q}$ and $\textbf{S}_j$ is the spin vector at site $j$. The magnetic moment at site $j$ in Bohr magneton is $g_jS_j$. The term $\frac{\gamma r_0}{2}$ in (10) containing the classical electron radius ($r_0$) and gyromagnetic ratio ($\gamma$) acts as an effective scattering length of per Bohr magneton, which is 2.695 fm \cite{Shirane2002}.
	
	\begin{figure}[t!]
		\centering{\includegraphics[clip,width=16cm]{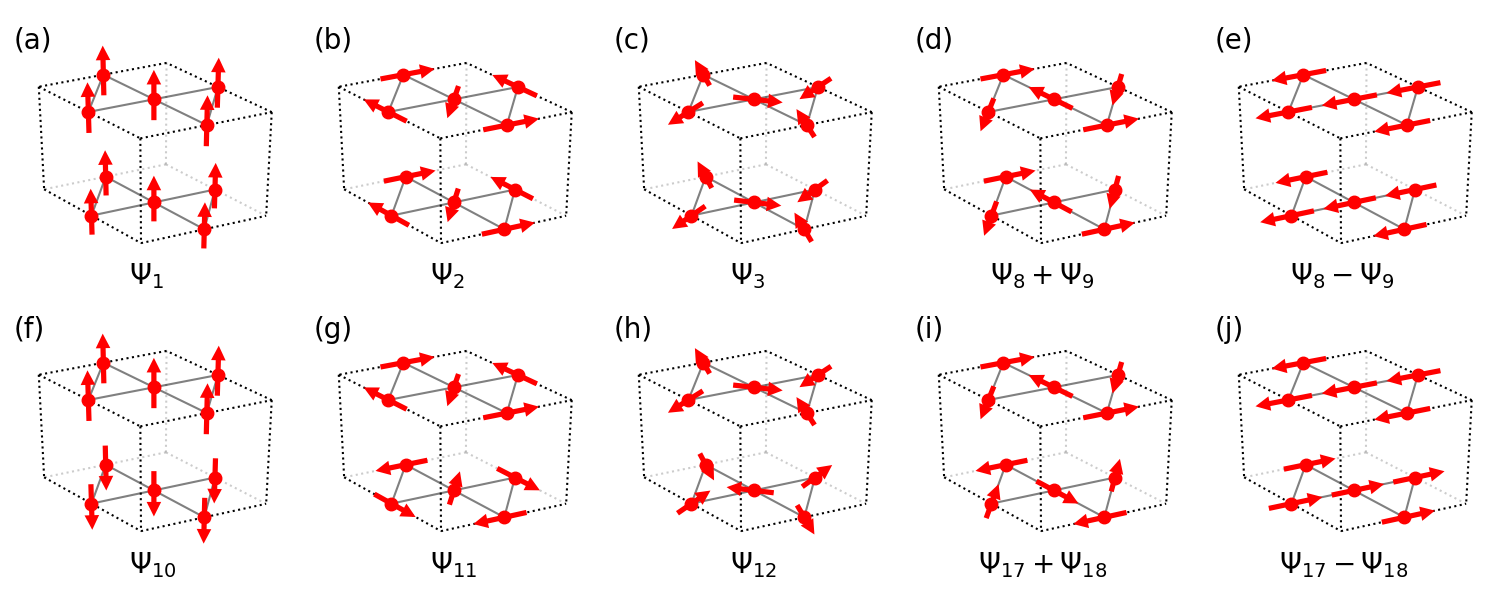}}
		\caption{Possible magnetic structures of the Fe kagome lattice based on the representation analysis.}
		\label{figs17}
	\end{figure}
	
	\begin{figure}[t!]
		\centering{\includegraphics[clip,width=11.5cm]{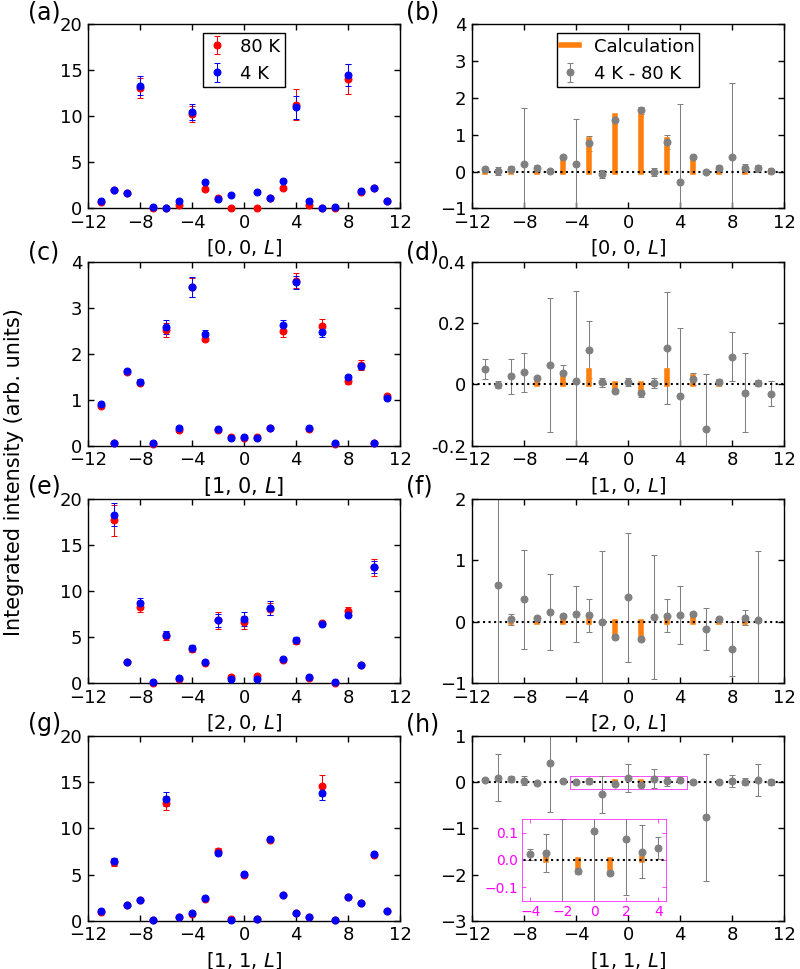}}
		\caption{(a), (c), (e), and (g) Integrated intensities of Bragg peaks along [0, 0, $L$], [1, 0, $L$], [2, 0, $L$], and [1, 1, $L$], respectively, at 80 K and 4 K. (b), (d), (f), and (h) The corresponding differences of the integrated intensities between 4 K and 80 K. The vertical orange bars indicate the calculated intensities from the determined magnetic structure. Inset of (h) is a zoom-in view of the data in the magenta box.}
		\label{figs18}
	\end{figure}
	
	For magnetic structures with the spins lying in the kagome plane, their magnetic structure factor at (0, 0, 1) can be explicitly written as
	\begin{equation}
		\textbf{F}_{\rm{M}}(\textbf{Q})=\frac{\gamma r_0}{2} gf(Q)\left[(\textbf{S}_1+\textbf{S}_2+\textbf{S}_3)e^{i2\pi z_{\rm{Fe}}}+(\textbf{S}_4+\textbf{S}_5+\textbf{S}_6)e^{-i2\pi z_{\rm{Fe}}}\right],
	\end{equation}
	where we have assumed the same $g$-factor and magnetic form factor for all Fe ions and taken $e^{-W_j} \approx 1$. $\textbf{S}_1$, $\textbf{S}_2$, and $\textbf{S}_3$ are the three spins in the first layer of the unit cell, and $\textbf{S}_4$, $\textbf{S}_5$, and $\textbf{S}_6$ are the three spins in the second layer (see Fig. \ref{figs17}). According to equation (12), we can immediately see that the magnetic structure factor at (0, 0, 1) vanishes for $\Psi_2$, $\Psi_3$, $\Psi_8+\Psi_9$, $\Psi_{11}$, $\Psi_{12}$, and $\Psi_{17}+\Psi_{18}$ [Fig. \ref{figs17}(b)-(d) and (g)-(i)], where $\textbf{S}_1+\textbf{S}_2+\textbf{S}_3=\textbf{0}$ and $\textbf{S}_4+\textbf{S}_5+\textbf{S}_6=\textbf{0}$ establish. On the contrary, our neutron diffraction experiment shows that significant magnetic scattering emerges at (0, 0, 1) below $T_{\rm{SR}}$ (see Fig. \ref{figs15}, Fig. \ref{figs16}, and Fig. 3 of the main text). Such observation excludes these 6 configurations and makes $\Psi_{17}-\Psi_{18}$ [Fig. \ref{figs17}(j)] the only possible magnetic structure below $T_{\rm{SR}}$.
	
	Based on the magnetic structure $\Psi_{17}-\Psi_{18}$, it is straightforward to calculate its magnetic scattering intensities and compare them with the experiment. Fig. \ref{figs18}(a), (c), (e), and (g) show the integrated intensities of a series of Bragg peaks at 80 K and 4 K. The data were taken at the CORELLI diffractometer, which can cover a large reciprocal space based on the Laue method. The difference of the integrated intensities between the two temperatures is shown in Fig. \ref{figs18}(b), (d), (f), and (h), which can be satisfactorily fitted with the magnetic structure $\Psi_{17}-\Psi_{18}$ [Fig. \ref{figs17}(j)]. The resultant magnetic moment per Fe ion is
	\begin{equation}
		M_{\rm{Fe}}=gS\mu_{\rm{B}}=1.51(3) \mu_{\rm{B}}.
	\end{equation}
	This moment size is consistent with previous reports \cite{VenturiniJAlloysCompd1992,MazetJPCM2000,CadoganJPCM2009}.
	
	\subsection{Possible non-collinear magnetic structure}
	
	Non-collinear magnetic structures have been observed in many kagome magnets. For instance, the B35 structure FeGe exhibits a non-collinear ``double-cone'' structure \cite{TengNature2022}. Notably, this compound can be regarded as the parent material of YbFe$_6$Ge$_6$. So it would be useful to estimate the upper bound of a possible non-collinear magnetic structure for YbFe$_6$Ge$_6$.
	
\begin{figure}[t!]
		\centering{\includegraphics[clip,width=12cm]{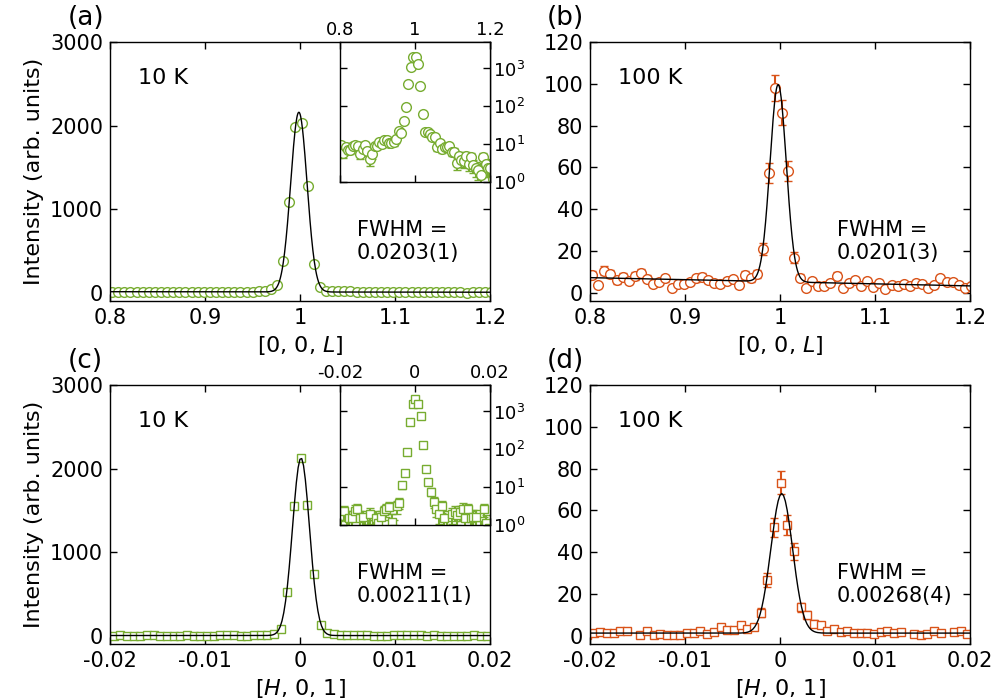}}
		\caption{Diffraction profiles along the [0, 0, $L$] and [$H$, 0, 1] directions at 10 K and 100 K, which are obtained from the data shown in Fig. 2(a) and (b) of the main text. Solid curves are the fits with Gaussian profiles. Insets of (a) and (c) show the same data with the intensity in log scale.}
		\label{figs19}
	\end{figure}

	Fig. \ref{figs19} presents the diffraction profiles around (0, 0, 1) along the [0, 0, $L$] and [$H$, 0, 1] directions, from which we can confirm the absence of additional peaks below $T_{\rm{SR}}$. Any satellite peak, if it exists, should be 3$\sim$4 orders of magnitude weaker than the major magnetic Bragg peak (i.e., in the level of the background). For comparison, the incommensurate magnetic peaks in FeGe (which are indicators of non-collinear magnetic structure) are about one order of magnitude weaker than the major magnetic peaks \cite{TengNature2022}.
	
\begin{figure}[h!]
		\centering{\includegraphics[clip,width=6cm]{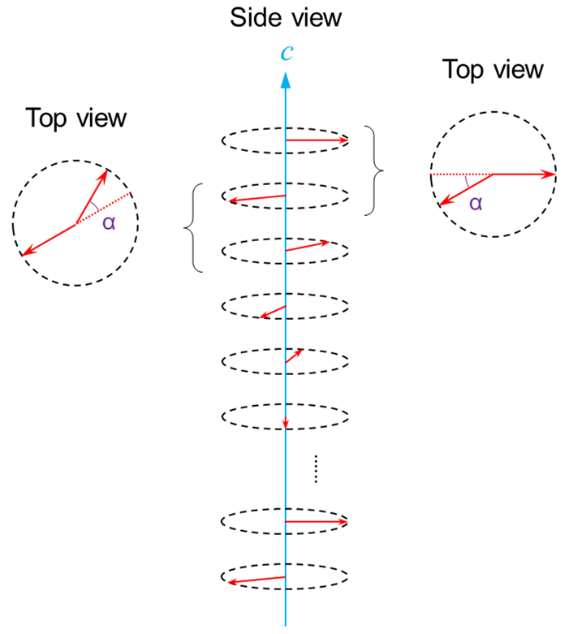}}
		\caption{Schematic of a $c$-axis helical structure supposed on the planar $A$-type antiferromagnetic structure. Each arrow represents a spin perpendicular to the $c$-axis. Two neighboring spins span a turn angle $\alpha$ + 180$^{\circ}$.}
		\label{figs20}
	\end{figure}

	To quantitatively estimate the upper bound of a possible non-colinear magnetic structure, we first assume the non-collinearity is a helical type \cite{JohnstonPRL2012,JohnstonPRB2017} - the spin rotates a small angle with respect to its neighbors. Fig. \ref{figs20} schematically shows a $c$-axis helical structure supposed on the $A$-type antiferromagnetic structure, where the spins are perpendicular to the $c$-axis. This structure will result in a satellite magnetic Bragg peak at \textbf{Q} + (0, 0, $\delta$), with $\delta$ = 2/(2$\pi$/$\alpha$). $\alpha$ + 180$^{\circ}$ is the turn angle spanned by two neighboring spins along the helical axis (i.e., the $c$-axis), where $\alpha$ directly quantifies the non-collinearity of two spins.
	
	Although we do not observe separate satellite peaks in YbFe$_6$Ge$_6$, they could be very close to integer-index positions so that cannot be resolved from the major peak. At 10 K, the full-width-at-half-maximum (FWHM) of the (0, 0, 1) peak is 0.0203(1) (in reciprocal lattice unit) along the [0, 0, $L$] direction [Fig. \ref{figs19}(a)]. That means for a possible satellite peak, the length of its propagation wave vector $\delta$ should be smaller than 0.0203. Therefore, the angle $\alpha$ is smaller than 3.65$^{\circ}$. Similarly, based on the FWHM along the [$H$, 0, 1] direction [Fig. \ref{figs19}(c)], we can estimate the upper bound of $\alpha$ to be 0.76$^{\circ}$ for spins in this direction. In the case of FeGe, an in-plane helical component is superposed on the axial $A$-type antiferromagnetic structure, which gives the ``double-cone'' structure \cite{Bernhard1984}. The turn angle $\alpha$ + 180$^{\circ}$ for the in-plane helical component is 194.4$^{\circ}$ and $\alpha$ = 14.4$^{\circ}$ \cite{Bernhard1984}.
	
	\section{Possible field-induced magnetic structure and small-angle neutron scattering}
	
	It has been widely recognized that field-induced static spin structures may give rise to AHE (or topological Hall effect). These structures usually have incommensurate propagation wave vectors of relatively small size, with the magnetic skyrmion as the most well-known example. To test whether this is the case for the observed AHE in YbFe$_6$Ge$_6$, we first estimate the real-space periodicity of the possible field-induced magnetic structure. Under this scenario, the anomalous Hall resistivity is proportional to the emergent magnetic field \cite{NeubauerPRL2009,TokuraChemRev2021}
	\begin{equation}
		\Delta \rho_{\rm{yx}} = PR_0b,
	\end{equation}
	where $P$ is the spin polarization, $R_0$ is the ordinary Hall coefficient, and $b$ is the emergent magnetic field. Spin polarization can be obtained from the magnetization data \cite{NeubauerPRL2009}
	\begin{equation}
		P = \frac{M}{M_{\rm{sat}}},
	\end{equation}
	where $M_{\rm{sat}}$ is the saturated magnetic moment. $R_0$ can be estimated from the fitted result of the ordinary Hall contribution (see Section II). On the other hand, the emergent magnetic field $b$ is related to the real-space periodicity of the magnetic structure with the following formula \cite{TokuraChemRev2021}
	\begin{equation}
		|b| \approx \frac{\phi_0}{\lambda^2},
	\end{equation}
	where $\phi_0 = \frac{h}{e}$ is the flux quantum and $\lambda$  is the periodicity of the magnetic structure in real space. By combining (14), (15), and (16), we can get
	\begin{equation}
		\lambda \approx \sqrt{\frac{\phi_0MR_0}{M_{\rm{sat}}\Delta \rho_{\rm{yx}}}}.
	\end{equation}
	For a magnetic field of 3 T, around which the AHE is most significant, we estimate $\lambda \approx 200 \rm{\AA}$. This real-space periodicity corresponds to a propagation wave vector in the order of $q \sim 0.03 \rm{\AA^{-1}}$ in reciprocal space. The emergent magnetic field $b \approx 12$ T.
	
	We use small-angle neutron scattering (SANS) to explore possible field-induced magnetic structures with such real-space periodicity. Our SANS experiment was performed with the QUOKKA instrument at the Australian Nuclear Science and Technology Organisation \cite{WoodJApplCryst2018}. The sample is a 15 mg YbFe$_6$Ge$_6$ single crystal. We attached Cd plates with a hole of $\sim$2.5 mm diameter on the sample holder to reduce scattering from the background. The sample holder was loaded into a horizontal field superconducting magnet. The direction of the magnetic field was parallel to the $\textbf{a}^\ast$ direction. Incident neutron beam with a wavelength of 5 $\rm{\AA}$ was parallel to the $c$-axis. We used low-$q$ and high-$q$ setups to cover momentum transfers from 0.008 $\rm{\AA}^{-1}$ to 0.04 $\rm{\AA}^{-1}$ and from 0.045 $\rm{\AA}^{-1}$ to 0.146 $\rm{\AA}^{-1}$, respectively.
	
\begin{figure}[t!]
		\centering{\includegraphics[clip,width=16cm]{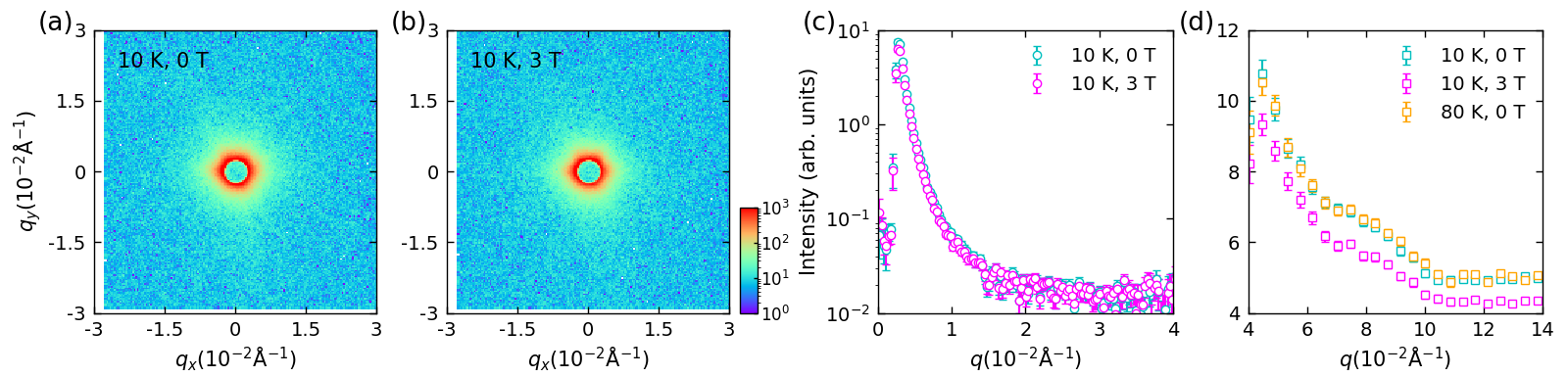}}
		\caption{(a) and (b) SANS patterns at 10 K, 0 T and 10 K, 3 T measured with the low-$q$ setup. (c) and (d) Azimuthally averaged SANS intensities as a function of momentum transfer in the low-$q$ and high-$q$ setups, respectively.}
		\label{figs21}
	\end{figure}

	Fig. \ref{figs21}(a) and (b) present the SANS patterns at 10 K, 0 T and 10 K, 3 T, respectively,  where we do not see Bragg peak in the range from 0.008 $\rm{\AA}^{-1}$ to 0.03 $\rm{\AA}^{-1}$ and the two patterns do not show a significant difference. These observations can be confirmed from the azimuthally averaged SANS intensities as a function of momentum transfer from 0.008 $\rm{\AA}^{-1}$ to 0.04 $\rm{\AA}^{-1}$ [Fig. \ref{figs21}(c)]. Moreover, with the high-$q$ setup, there is still no evident difference for the data at 10 K, 0 T and 10 K, 3 T [Fig. \ref{figs21}(d)]. Therefore, we can rule out field-induced magnetic Bragg peaks in the momentum transfer from 0.008 $\rm{\AA}^{-1}$ to 0.14 $\rm{\AA}^{-1}$.
	
	As discussed above, if the observed AHE is due to field-induced static spin texture, the corresponding propagation wave vector is expected to be in the order of $q \sim 0.03 \rm{\AA^{-1}}$. So, the absence of field-induced peaks from 0.008 $\rm{\AA}^{-1}$ to 0.14 $\rm{\AA}^{-1}$ makes it very unlikely that the AHE is due to field-induced static spin textures.
	
	\section{Inelastic neutron scattering}
	
	\begin{figure}[b!]
		\centering{\includegraphics[clip,width=15cm]{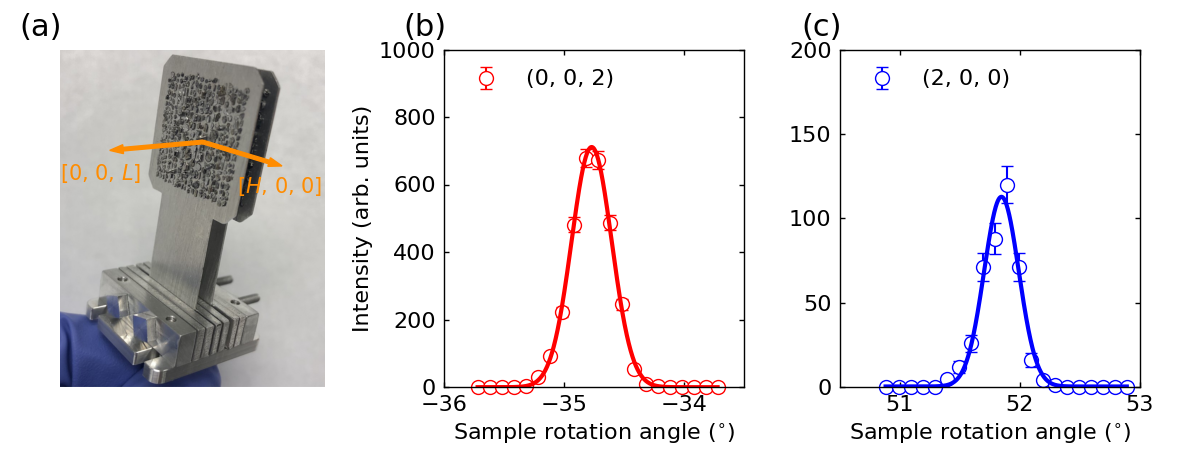}}
		\caption{(a) Coaligned single crystals used in our INS experiments. (b) and (c) Rocking scans performed at Bragg peaks (0, 0, 2) and (2, 0, 0), respectively. Solid lines are the fits with Gaussian profiles.}
		\label{figs22}
	\end{figure}
	
	Inelastic neutron scattering (INS) experiments with time-of-flight neutron spectrometers were performed on a coaligned single crystal array with the ARCS spectrometer at Spallation Neutron Source, Oak Ridge National Laboratory \cite{Abernathy2012} and the MERLIN spectrometer at ISIS Spallation Neutron Source, the Rutherford Appleton Laboratory \cite{BewleyPhysicaB2006}. The obtained time-of-flight neutron data was reduced and analyzed with Mantid \cite{Arnold2014} and HORACE \cite{Ewings2016}. An additional INS experiment was performed on the same sample with the cold neutron spectrometer HODACA at Japan Research Reactor-3 (JRR3) \cite{KikuchiJPSJ2024}. The coaligned single crystal array is shown in Fig. \ref{figs22}(a), which consists of about 1000 pieces ($\sim$0.9 gram) of small single crystals glued on two aluminum plates. The samples on the plates were oriented such that the scattering plane was ($H$, 0, $L$). For the time-of-flight neutron scattering data shown in the presented figures, their momentum- and/or energy-integrated ranges are listed in Table \ref{tb5}.
	
\begin{figure}[t!]
		\centering{\includegraphics[clip,width=13cm]{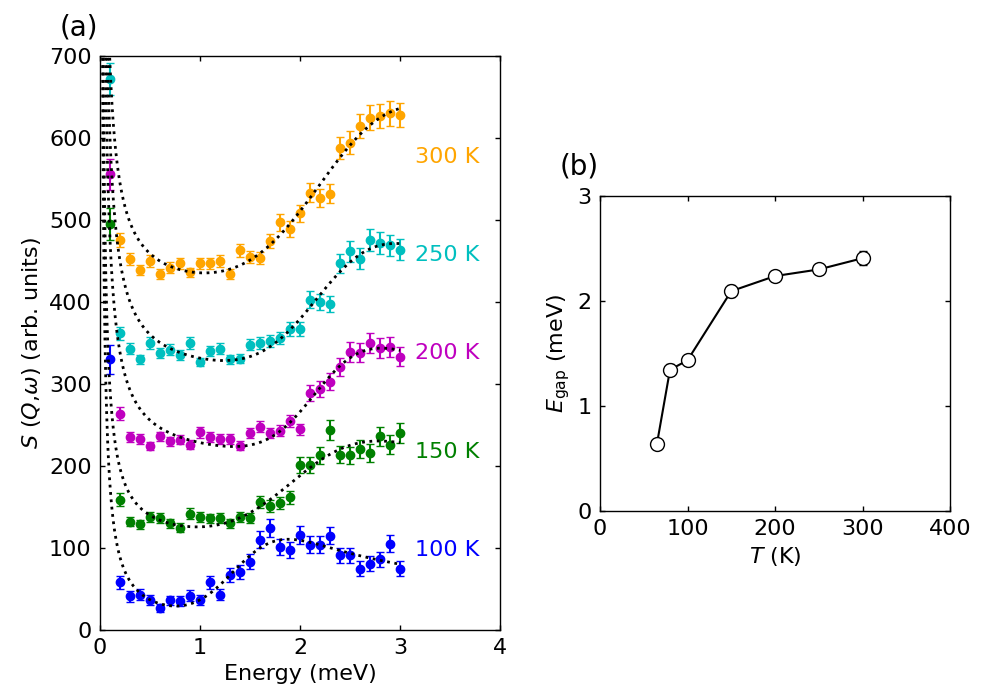}}
		\caption{(a) Low-energy spin excitations around (0, 0, 1) at selected temperatures above 100 K. Data are evenly offset for clarity. Dashed curves are the fits as described in the text. (b) Temperature dependence of the fitted gap size. The fitting errorbar is smaller than the symbol size.}
		\label{figs23}
	\end{figure}

	The data shown in Fig. 4(a) and (b) of the main text were measured at the HODACA spectrometer, which was run in the triple-axis mode. Rocking scans performed at Bragg peaks (0, 0, 2) and (2, 0, 0) indicate a sample mosaic spread of 0.40(1)$^\circ$ and 0.34(2)$^\circ$, respectively [Fig. \ref{figs22}(b) and (c)], which are estimated by their FWHMs. For INS measurements, constant energy scans at fixed momentum positions were performed with a fixed final neutron energy of $E_{f}$ = 3.636 meV. The energy resolution is about 0.14 meV at zero energy transfer. The typical counting time for each energy position is about 5 minutes. Additional INS data taken above 100 K are shown in Fig. \ref{figs23}(a), which supplement the data presented in Fig. 4(a) of the main text. To extract the energy gap size, we fit these data with an error function multiplied by the Bose factor \cite{ChenNC2024}
	\begin{equation}
		I (E, T)=\left[I_0+A\cdot \rm{erf}\left(\frac{\it{E}-\it{E}_{\rm{gap}}}{\sigma}\right)\right]\cdot \frac{1}{1-e^{-E/k_{\rm{B}}T}},
	\end{equation}
	where erf($x$) represents the error function, $I_0$, $A$, $\sigma$, and $E_{\rm{gap}}$ (the gap size) are fitting parameters. The fitting results are shown with dashed curves in Fig. 4(a) of the main text and Fig. \ref{figs23}(a). The obtained gap sizes at various temperatures are presented in Fig. \ref{figs23}(b). From 65 K to 300 K, we can see that the spin-wave gap size gradually increases, which is consistent with the increase of the spin-flop field above $T_{\rm{SR}}$ [see the phase diagram in Fig. \ref{figs5}(c)] \cite{AvilaJPCM2005}.
	
\begin{figure}[h!]
		\centering{\includegraphics[clip,width=13cm]{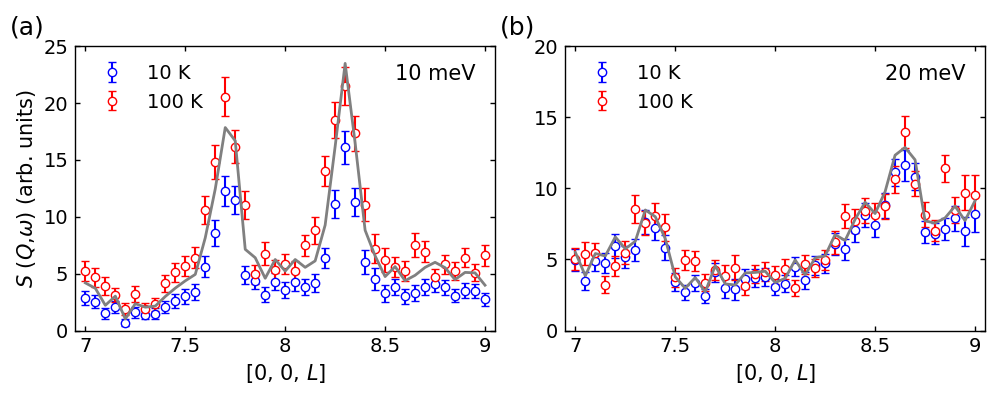}}
		\caption{(a) and (b) Constant energy cuts around (0, 0, 8) along the [0, 0, $L$] direction at 10 meV and 20 meV, respectively. Grey curves represent the calculated intensities at 100 K by correcting the 10 K data with the Bose factor.}
		\label{figs24}
	\end{figure}

	The data shown in Fig. 5(a)-(d) of the main text were measured at the ARCS spectrometer with an incident neutron energy ($E_{\rm{i}}$) of 40 meV. The energy resolution is about 2.4 meV at zero energy transfer. The sample was rotated along the [-0.5$K$, $K$, 0] direction for about 150$^{\circ}$ in 1$^{\circ}$ per step. The counting time for each sample rotation angle is about 6 minutes. To reveal the different dependence behaviors of the phonon and magnon, constant energy cuts around (0, 0, 8) along the [0, 0, $L$] direction are presented in Fig. \ref{figs15}. These signals mainly come from phonons due to the large momentum transfer. We can see that the intensity change between 10 K and 100 K follows the Bose factor, which is in contrast to the magnon signals shown in Fig. 5(c) and (d) of the main text. Fig. \ref{figs24} presents more INS data of 10 K and 100 K measured with $E_{\rm{i}}$ = 100 meV, from which we can see the higher energy part of the magnon emanating from (0, 0, 3). The spin-wave bandwidth is about 40 meV for the dispersion perpendicular to the kagome plane, and is larger than 60 meV along the kagome plane. In these measurements, the energy resolution is about 8 meV at zero energy transfer.
	
	The data shown in Fig. 5(e)-(i) of the main text were measured at the MERLIN spectrometer with $E_{\rm{i}}$ = 15 meV. The energy resolution is about 0.7 meV at zero energy transfer. Since the low-energy magnetic excitations concentrate very close to the Brillouin zone center, we rotated the sample along the [-0.5$K$, $K$, 0] direction for about 60$^{\circ}$ in 1$^{\circ}$ per step. The measured reciprocal space covers a region along the [0, 0, $L$] direction. The counting time for each sample rotation angle is about 14 minutes. Fig. \ref{figs26} presents the distribution of the low-energy spin excitations in the hexagonal Brillouin zone, which confirms the rod-like feature shown in Fig. 5(e)-(i) of the main text.
	
	\begin{figure}[h!]
		\centering{\includegraphics[clip,width=13cm]{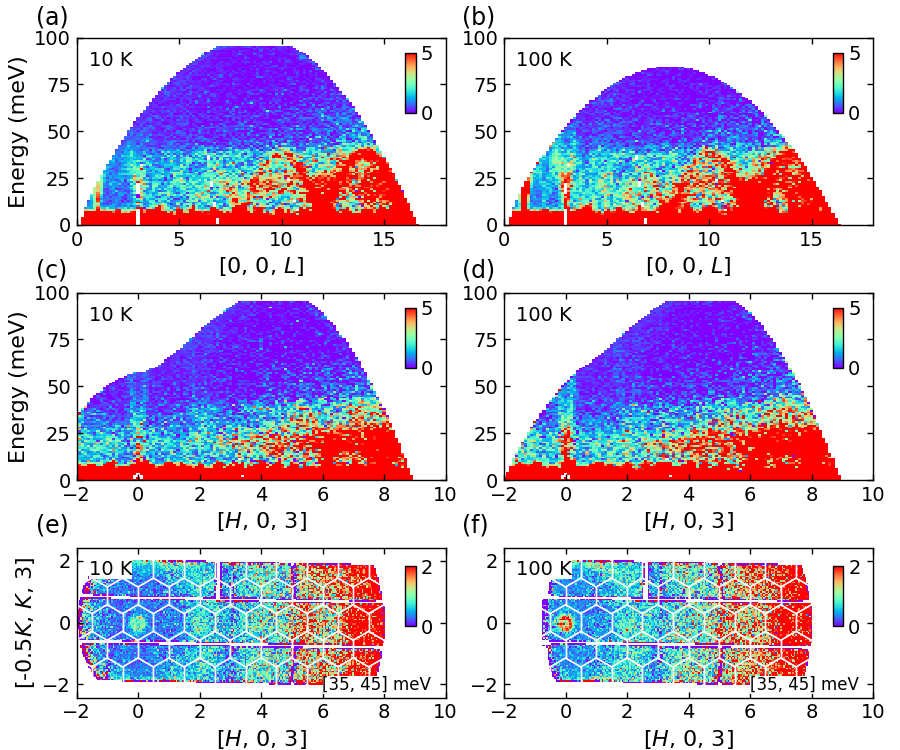}}
		\caption{(a) and (b) Excitation spectra along [0, 0, $L$] at 10 K and 100 K, respectively. (c) and (d) Excitation spectra along [$H$, 0, 3] at 10 K and 100 K, respectively. (e) and (f) Constant energy cuts around 40 meV of the ($H$, $K$, 3) plane at 10 K and 100 K, respectively. White hexagons are the Brillouin zone boundaries.}
		\label{figs25}
	\end{figure}
	
	\begin{figure}[h!]
		\centering{\includegraphics[clip,width=16cm]{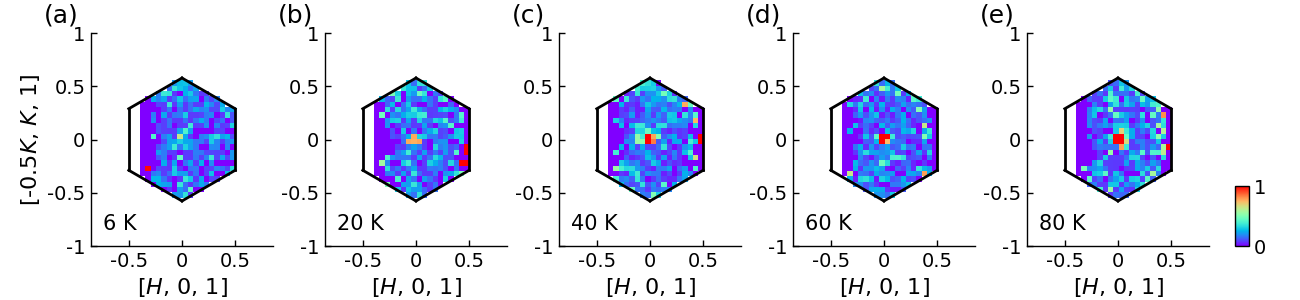}}
		\caption{(a)-(e) Constant energy cuts of the low-energy spin excitations at selected temperatures. Data are obtained by integrating the energy from 1 meV to 3 meV. Only the data of the first hexagonal Brillouin zone centering at (0, 0, 1) are shown.}
		\label{figs26}
	\end{figure}
	
	\begin{table}[h]
		\caption{
			Crystal data and structure refinement for YbFe$_6$Ge$_6$.
		}
		\begin{ruledtabular}
			\begin{tabular}{cc}
				Empirical formula&YbFe$_6$Ge$_6$\\
				Formula weight& 943.66\\
				Temperature&100.00(10) K\\
				Wavelength&0.71073 \AA\\
				Crystal system&Hexagonal\\
				Space group&$P$6/$mmm$\\
				Unit cell dimensions&a = b = 5.0795(3) \AA, c = 8.0806(9) \AA\\
				&$\alpha$ = $\beta$ = 90$^{\circ}$, $\gamma$ = 120$^{\circ}$\\
				Volume&180.56(2)\AA$^3$\\
				Z&1\\
				Density (calculated)&8.6070 g$\cdot$cm$^{-3}$\\
				Absorption coefficient&48.511 mm$^{-1}$\\
				F(000)&418\\
				Crystal size&0.11 $\times$ 0.10 $\times$ 0.07 mm$^3$\\
				Theta range for data collection&4.616$^{\circ}$ to 47.565$^{\circ}$\\
				Index ranges&-10 $\le$ $H$ $\le$ 7, -8 $\le$ $K$ $\le$ 8, -12 $\le$ $L$ $\le$ 10\\
				Reflections collected&1449\\
				Independent reflections&310 [R(int) = 0.0279]\\
				Completeness to theta = 25.242$^{\circ}$&100.0 \%\\
				Refinement method&Full-matrix least-squares on F$^2$\\
				Data / restraints / parameters&310 / 0 / 9\\
				Goodness-of-fit on F$^2$&2.06\\
				Final R indices [I $>$ 3sigma(I)]&R1 = 0.0298, wR2 = 0.0749\\
				R indices (all data)&R1 = 0.0320, wR2 = 0.0755\\
				Extinction coefficient&0.0160(19)\\
			\end{tabular}	
		\end{ruledtabular}
		\label{tb1}
	\end{table}
	
	\begin{table}[h]
		\caption{Atomic coordinates and equivalent isotropic displacement parameters of YbFe$_6$Ge$_6$ at 100 K.}
		\begin{ruledtabular}
			\begin{tabular}{ccccccc}
				\textbf{Atom}&\textbf{Wyckoff.}&\textbf{Occ.}&\textbf{x}&\textbf{y}&\textbf{z}&\textbf{U$\rm_{eq}$}\\
				\midrule
				Yb1&1$a$&1&0&0&0&0.0030(2)\\
				Ge1&2$c$&1&0.3333&0.6667&0&0.0066(2)\\
				Ge2&2$d$&1&0.3333&0.6667&0.5&0.0042(2)\\
				Ge3&2$e$&1&0&0&0.3456(2)&0.0030(2)\\
				Fe1&6$i$&1&0.5&0&0.2491(1)&0.0032(2)\\
			\end{tabular}	
		\end{ruledtabular}
		\label{tb2}
	\end{table}
	
	\begin{table}[h]
		\caption{Elemental analysis report from the spectrum in Fig. \ref{figs4}.}
		\begin{ruledtabular}
			\begin{tabular}{cccc}
				Element&Weight percentage (\%)&Weight percentage error (\%)&Atomic percentage (\%)\\
				\midrule
				Yb&18.51&0.83&7.75\\
				Fe&36.31&0.73&47.13\\
				Ge&45.18&0.73&45.12\\
				\midrule
				Total&100&-&100\\
			\end{tabular}	
		\end{ruledtabular}
		\label{tb3}
	\end{table}
	
	\begin{table}[b]
		\caption{Irreducible representations (IRs) and basis vectors (BVs) for the space group $P$6/$mmm$ with magnetic propagation vector $\textbf{k}_{\rm{m}}$ = (0, 0, 0).}
		\begin{ruledtabular}
			\begin{tabular}{cccccccc}
				IR&BV&Fe1&Fe2&Fe3&Fe4&Fe5&Fe6\\
				\midrule  
				$\Gamma_2$&$\Psi_1$&(0,0,1)&(0,0,1)&(0,0,1)&(0,0,1)&(0,0,1)&(0,0,1)\\
				\midrule
				$\Gamma_3$&$\Psi_2$&(1,0,0)&(0,1,0)&(-1,-1,0)&(1,0,0)&(0,1,0)&(-1,-1,0)\\
				\midrule
				$\Gamma_4$&$\Psi_3$&(0.5,1,0)&(-1,-0.5,0)&(0.5,-0.5,0)&(0.5,1,0)&(-1,-0.5,0)&(0.5,-0.5,0)\\
				\midrule
				\multirow{2}{*}{$\Gamma_5$}&$\Psi_4$&(0,0,0)&(0,0,-1)&(0,0,1)&(0,0,0)&(0,0,-1)&(0,0,1)\\
				&$\Psi_5$&(0,0,1)&(0,0,-0.5)&(0,0,-0.5)&(0,0,1)&(0,0,-0.5)&(0,0,-0.5)\\
				\midrule
				\multirow{4}{*}{$\Gamma_6$}&$\Psi_6$&(0.5,1,0)&(0.5,0.25,0)&(-0.25,0.25,0)&(0.5,1,0)&(0.5,0.25,0)&(-0.25,0.25,0)\\
				&$\Psi_7$&(0,0,0)&(0,-1,0)&(-1,-1,0)&(0,0,0)&(0,-1,0)&(-1,-1,0)\\
				&$\Psi_8$&(0,0,0)&(-1,-0.5,0)&(-0.5,0.5,0)&(0,0,0)&(-1,-0.5,0)&(-0.5,0.5,0)\\
				&$\Psi_9$&(1,0,0)&(0,-0.5,0)&(0.5,0.5,0)&(1,0,0)&(0,-0.5,0)&(0.5,0.5,0)\\
				\midrule
				$\Gamma_7$&$\Psi_{10}$&(0,0,1)&(0,0,1)&(0,0,1)&(0,0,-1)&(0,0,-1)&(0,0,-1)\\
				\midrule
				$\Gamma_{9}$&$\Psi_{11}$&(0.5,1,0)&(-1,-0.5,0)&(0.5,-0.5,0)&(-0.5,-1,0)&(1,0.5,0)&(-0.5,0.5,0)\\
				\midrule
				$\Gamma_{10}$&$\Psi_{12}$&(1,0,0)&(0,1,0)&(-1,-1,0)&(-1,0,0)&(0,-1,0)&(1,1,0)\\
				\midrule
				\multirow{2}{*}{$\Gamma_{11}$}&$\Psi_{13}$&(0,0,0.5)&(0,0,-1)&(0,0,0.5)&(0,0,-0.5)&(0,0,1)&(0,0,-0.5)\\
				&$\Psi_{14}$&(0,0,-1)&(0,0,0)&(0,0,1)&(0,0,1)&(0,0,0)&(0,0,-1)\\
				\midrule
				\multirow{4}{*}{$\Gamma_{12}$}&$\Psi_{15}$&(1,0,0)&(0,-0.5,0)&(0.5,0.5,0)&(-1,0,0)&(0,0.5,0)&(-0.5,-0.5,0)\\
				&$\Psi_{16}$&(0,0,0)&(1,0.5,0)&(0.5,-0.5,0)&(0,0,0)&(-1,-0.5,0)&(-0.5,0.5,0)\\
				&$\Psi_{17}$&(0,0,0)&(0,1,0)&(1,1,0)&(0,0,0)&(0,-1,0)&(-1,-1,0)\\
				&$\Psi_{18}$&(0.5,1,0)&(0.5,0.25,0)&(-0.25,0.25,0)&(-0.5,-1,0)&(-0.5,-0.25,0)&(0.25,-0.25,0)\\
				
			\end{tabular}	
		\end{ruledtabular}
		\label{tb4}
	\end{table}

	\begin{table}[b]
		\caption{Momentum- and/or energy-integrated ranges for the INS data presented in figures.}
		\begin{ruledtabular}
			\begin{tabular}{cccc}
				Data &\makecell{Momentum-integrated\\range (r.l.u.)}& \makecell{Energy-integrated\\range (meV)}& Spectrometer\\
				\midrule
				Fig. 5(a) and (b)&\makecell{[-0.1, 0.1] along [$H$, 0, 0] \\ $$[-0.1, 0.1] along [-0.5$K$, -$K$, 0]}&-&ARCS\\
				\hline
				Fig. 5(c) and (d)&\makecell{[-0.1, 0.1] along [$H$, 0, 0] \\ $$[-0.1, 0.1] along [-0.5$K$, -$K$, 0]}&[-1, 1]&ARCS\\
				\hline
				Fig. 5(e)-(i)&\makecell{[-0.1, 0.1] along [$H$, 0, 0] \\ $$[-0.1, 0.1] along [-0.5$K$, -$K$, 0]}&-&MERLIN\\
				\hline
				Fig. \ref{figs24}&\makecell{[-0.1, 0.1] along [$H$, 0, 0] \\ $$[-0.1, 0.1] along [-0.5$K$, -$K$, 0]}&[-1, 1]&ARCS\\
				\hline
				Fig. \ref{figs25}(a)-(d)&\makecell{[-0.1, 0.1] along [$H$, 0, 0] \\ $$[-0.1, 0.1] along [-0.5$K$, -$K$, 0]}&-&ARCS\\
				\hline
				Fig. \ref{figs25}(e) and (f)&\makecell{[-0.2, 0.2] along $[0, 0, L]$}&[-5, 5]&ARCS\\
				\hline
				Fig. \ref{figs26}&\makecell{[-0.15, 0.15] along $[0, 0, L]$}&[-1, 1]&MERLIN\\
				
			\end{tabular}	
		\end{ruledtabular}
		\label{tb5}
	\end{table}

	\clearpage
	%\nocite{apsrev42Control}
	%\bibliographystyle{apsrev4-2}
	%\bibliography{yfg_reference}

\end{document}